\documentclass[runningheads,a4paper]{llncs}

\usepackage[utf8]{inputenc}
\usepackage{booktabs} 
\usepackage{graphicx}
\usepackage{units}
\usepackage{blindtext}
\usepackage{mathtools}
\usepackage{cite}
\usepackage{xcolor,xspace}
\usepackage{amsmath}
\usepackage{paralist}
\usepackage{subfig}
\usepackage{amsmath}
\usepackage[title]{appendix}
\usepackage{tikz}
\usepackage{booktabs}
\usepackage{pifont}
\usepackage{multirow}
\usepackage{hhline}
\usepackage{balance}
\usepackage{footnote}
\usepackage{arydshln}
\usepackage{enumitem}
\usepackage[flushleft]{threeparttable}
\usetikzlibrary{calc}
\usetikzlibrary{decorations.pathreplacing}

\usepackage{listings}
\usepackage{algorithm}
\usepackage{algpseudocode}

\usepackage{xcolor}
\usepackage{listings}
\usepackage{lstlinebgrd}

\newtheorem{remarkk}{Remark}
\newtheorem{observation}{Observation}
\definecolor{mGreen}{rgb}{0,0.6,0}
\definecolor{mGray}{rgb}{0.5,0.5,0.5}
\definecolor{lgreen}{RGB}{144,238,144}
\definecolor{mPurple}{rgb}{0.58,0,0.82}
\definecolor{backgroundColour}{rgb}{0.95,0.95,0.92}
\definecolor{Blue}{RGB}{42, 0, 255}
\definecolor{Red}{RGB}{178,34,34}
\definecolor{Orange}{RGB}{255,69,0}
\newcommand{\cmark}{\ding{51}}%
\newcommand{\xmark}{\ding{55}}%

\lstdefinestyle{ShStyle}{
    commentstyle=\color{mGreen},
    keywordstyle=\color{magenta},
    numberstyle=\tiny\color{mGray},
    stringstyle=\color{mPurple},
    basicstyle=\footnotesize\ttfamily,
    breakatwhitespace=false,         
    breaklines=true,                 
    captionpos=b,                    
    keepspaces=true,                 
    numbers=left,                    
    numbersep=5pt,                  
    showspaces=false,                
    showstringspaces=false,
    showtabs=false,                  
    tabsize=2,
    language=bash,
    frame=single
}

\lstdefinestyle{CStyle}{
    commentstyle=\color{mGreen},
    keywordstyle=\color{magenta},
    numberstyle=\tiny\color{mGray},
    stringstyle=\color{mPurple},
    basicstyle=\tiny,
    breakatwhitespace=false,         
    breaklines=true,                 
    captionpos=b,                    
    keepspaces=true,                 
    numbers=left,                    
    numbersep=5pt,                  
    showspaces=false,              
    showstringspaces=false,
    showtabs=false,                  
    tabsize=2,
    language=C,
    frame=single
}

\lstdefinelanguage
   [x64]{Assembler}     
   {morekeywords={push, mov, sub, add, call, jle, jne, cmp, cdqe, movzx, lea, movsxd}} 

\algrenewcommand\algorithmicrequire{\textbf{Input:}}
\algrenewcommand\algorithmicensure{\textbf{Output:}}

\newcommand{\acron}{{{\sf ALICE}}\xspace}

\newcommand\karim[1]{#1}
\newcommand\oak[1]{#1}
\long\def\ignore#1{}

\newcommand{\projectfullname} {Augmentation and Legacy-software Instrumentation of Cryptographic Executables~}
\newcommand{\projectname} {\acron}

\usepackage{expl3,xparse}

\ExplSyntaxOn
\NewDocumentCommand \lstcolorlines { O{green} m }
{
 \clist_if_in:nVT { #2 } { \the\value{lstnumber} }{ \color{#1} }
}
\ExplSyntaxOff

\begin{document}

\title{Towards Automated Augmentation and Instrumentation of Legacy Cryptographic Executables: Extended Version}

\author{Karim Eldefrawy\inst{1} \and Michael Locasto\inst{1} \and Norrathep Rattanavipanon\thanks{Work done partially while at SRI International.}\inst{2} \and Hassen Saidi\inst{1}}
\institute{SRI International\\\email{\{karim.eledefrawy, michael.locasto, hassen.saidi\}@sri.com}\\ \and Prince of Songkla University, Phuket Campus\\\email{norrathep.r@phuket.psu.ac.th}}

\maketitle

\begin{abstract}

Implementation flaws in cryptographic libraries, design flaws in algorithms underlying
cryptographic primitives, and weaknesses in protocols using both, can all lead to exploitable vulnerabilities in software. 
Manually fixing such issues is challenging and resource consuming, especially 
when maintaining legacy software that contains broken or outdated cryptography, and 
for which source code may not be available. 
\karim{While there is existing work on identifying cryptographic primitives (often in the context of malware analysis),
none of this prior work has focused on replacing such primitives with stronger (or more secure ones)
after they have been identified.}
This paper explores feasibility of designing and implementing a toolchain for \textit{\projectfullname (\projectname)}. 
The key features of \projectname are:
(i) automatically detecting and extracting implementations 
of weak or broken cryptographic primitives from binaries without requiring source code or debugging symbols,
(ii) identifying the context and 
scope in which  such primitives are used, and performing 
program analysis to determine the effects of replacing
such implementations with more secure ones, and
(iii) replacing implementations of weak primitives
with those of stronger or more secure ones. 
We demonstrate practical feasibility of our approach on cryptographic hash functions with several popular cryptographic libraries and real-world programs
of various levels of complexity. 
Our experimental results show that \acron can locate and replace insecure hash functions, even in large binaries (we tested ones of size up to 1.5MB),
while preserving existing functionality of the original binaries, and while incurring minimal execution-time overhead in the rewritten binaries. 
We also open source \acron's code at \texttt{https://github.com/SRI-CSL/ALICE}.

\end{abstract}

\section{Introduction}
Cryptography is instrumental to implementing security services such as confidentiality, integrity, and authenticity in most software (both, new and legacy). In practice, proper usage and correct implementation of cryptographic primitives are difficult; vulnerabilities often occur due to misuse or erroneous implementations of cryptographic primitives.
Example vulnerabilities arising from misuse of cryptography include weak and/or broken random number generators, resulting in enabling an adversary to recover servers' private keys~\cite{PsAndQs}. Cryptographic APIs are sometimes misused by software developers, e.g., causing applications to be insecure against specific attacks, such as chosen plaintext~\cite{egele2013empirical} which a typical software developer may be unaware of.

In addition, incorrect implementations of cryptographic primitives can result in leakage of secrets through side-channels~\cite{crypto-side-channel-1} or through ``dead memory''~\cite{khunt}.
Other vulnerabilities in software for embedded and generic systems include implementation flaws in cryptographic libraries (e.g., the
HeartBleed~\cite{HeartBleed} and Poodle~\cite{Poodle} vulnerabilities in the OpenSSL library), weaknesses in protocol suites (e.g., cryptographic weakness in HTTPS implementations~\cite{postcards, Logjam-Attack}), and algorithmic vulnerabilities in
cryptographic primitives (e.g., an unknown collision 
attack on the MD5 hash function~\cite{Counter-Cryptanalysis}
or a chosen-prefix collision attack on the SHA1 hash function~\cite{leurentsha}).
Even after such vulnerabilities are discovered, it may take a while before
appropriate fixes are applied to existing software as demonstrated by
a recent large-scale empirical study~\cite{li2017large} that showed
many software projects did not patch cryptography-related vulnerabilities
for a full year after their public disclosure.
This represents a large window for adversaries to exploit
such vulnerabilities.

\emph{We argue that (automated) tools that assist software and system designers, and developers, in performing identification, analysis, and replacement in binaries
(without requiring source code) can help shorten such vulnerability window, especially for legacy software.}
To address this issue, we explore feasibility of designing and developing a toolchain for \textit{\projectfullname (\projectname)}. 

~

\textbf{Contributions:} Specifically, our goal is to make the following contributions:
\begin{compactenum}

\item We design the  \acron framework to automatically augment and instrument executables with broken or insecure cryptographic primitives. We also open source \acron's code at \texttt{https://github.com/SRI-CSL/ALICE}.

\item We develop heuristics to identify (binary) code segments implementing cryptographic primitives (see Table~\ref{tab:detected-primitives} in Appendix~\ref{apdx:detect_prim} for a list of such primitives).

\item We develop heuristics to determine the scope of the (binary) code segments requiring augmentation if the cryptographic primitives are replaced with stronger ones. 

\item We implement \acron and experimentally evaluate its performance on several executable open source binaries of varying complexity. 


\end{compactenum}
~

\textbf{Outline:} The rest of this paper is organized as follows: Section \ref{sec:rel-work} discusses related work. 
Section \ref{sec:approach-overview} overviews the \acron toolchain, while Section \ref{sec:alice-details} contains 
its design details. Due to space constraints, implementation details are described in Appendix~\ref{sec:impl}.
Section \ref{sec:eval} contains the results of our experimental evaluations. Section~\ref{sec:limitation} discusses \acron 's limitations, while Section \ref{conclusion} concludes the paper.


\section{Related Work \label{sec:rel-work}}

\ignore{
\oak{The closest related work is a recent tool, called Fennec~\footnote{https://blog.trailofbits.com/2019/09/02/rewriting-functions-in-compiled-binaries/}, that can be used to replace 
insecure cryptographic functions in binaries.
In contrast to \acron, Fennec assumes: (1) locations of insecure functions are manually detected and supplied to the tool, and (2) the replacement function has the same function parameters as the insecure function.

Our work contain neither of those assumptions.}}


\noindent\textbf{Identifying Cryptographic Primitives.}
Several publicly available tools \cite{findcrypt, kanal, hcdetector} utilize static analysis to identify cryptographic primitives by detecting known (large) constants used in their operation.
Such constants, for example, can be in the form of look-up tables (e.g., S-Boxs in AES) or a fixed initialization vectors/values (e.g., IV in SHA-128/256).
Such tools do not always produce accurate results as the detected algorithm may be another function or another cryptographic primitive that uses the same constant values~\cite{LGF+15}.
They are also ineffective when dealing with obfuscated programs~\cite{CFM+12}.

In terms of academic efforts, Lutz~\cite{lutz2008towards} detects block ciphers from execution traces based on three heuristics: the presence of loops,
high entropy, and integer arithmetic. Grobert et al.~\cite{GWH+11} introduce an additional heuristic to extract cryptographic parameters from such execution traces and 
identify primitives by comparing the input-output relationships with those of known cryptographic functions.
Lestringant et al.~\cite{LGF+15} propose a static method based on data flow graph isomorphism
to identify symmetric cryptographic primitives. 
Recently, the CryptoHunt \cite{CryptoHunt17} tool develop a new technique
called bit-precise symbolic loop mapping to identify cryptographic
primitives in obfuscated binaries. 

Our work focuses on non-obfuscated programs as we target common (and possibly legacy) software and not malware.
We rely on finding known constants to identify cryptographic primitives as our first step.
We then improve the accuracy of detection by applying a heuristic
based on input-output relationships, similar to the work in~\cite{GWH+11}.
In contrast to~\cite{GWH+11}, our identification algorithm does not require program execution traces. 
 

%
\karim{While there is existing work on identifying executable segments implementing cryptographic primitives, 
	none of such work investigates the problem of replacing an identified weak primitive
	with a more secure one. Such replacement requires non-trivial operations, even if one can successfully identify executable segments implementing cryptographic primitives. To accomplish such replacement, one has to perform the following: (1) determining all changes throughout the binary necessary for replacing the identified primitive, and (2) rewriting the binary to apply all of the determined changes.
	\emph{To the best of our knowledge, there is no prior work addressing the first task, as a standalone, or in conjunction with the second.} The second task can be tackled using slight modifications of existing binary rewriting techniques.
	In this paper, we categorize different types of necessary changes one may require when replacing a cryptographic primitive, and then discuss how to locate and rewrite each category of such changes in Section~\ref{sec:scope}.
	In the rest of this section, we overview general binary rewriting techniques. 
}

\noindent\textbf{Rewriting Binaries.}
Binary rewriting is a technique that transforms a binary executable into another without requiring the original's source code.
Typically, the transformed binary must preserve the functionality of the original one while possibly augmenting it with extra functionalities. 
There are two main categories of binary rewriting: static and dynamic binary rewriting.

In \textit{static binary rewriting}~\cite{detours, bauman12superset, anand2013compiler, edwards2001vulcan}, the original binary is modified offline without executing it.
Static binary rewriting is typically performed by replacing the original instructions with an unconditional jump that redirects the program control flow to the rewritten instructions, stored in a different area of the binary. 
This relocation can be done at different levels of granularity such as inserting a jump for each modified instruction, for the entire section or for each routine containing modified instructions.
Static binary rewriting often requires disassembling the entire binary and thus incurs high overhead during the rewriting phase,
but typically results in small runtime overhead in the rewritten binary.
This technique is thus well-suited for scenarios where the runtime performance of the rewritten binary is a primary concern.
Another approach for static rewriting is to transform the binary into the relocatable disassembled code and directly rewrite instructions in the transformed code.
Doing so completely eliminates runtime and size overhead in the rewritten binary.
Nonetheless, this approach relies on many heuristics and assumptions for identifying and recovering all relocating symbols
and is still subject to the high overhead during the rewriting phase.
Some example tools that are based on this approach are \texttt{Uroboros}~\cite{uroboros} and \texttt{Ramblr}~\cite{ramblr}.

\textit{Dynamic binary rewriting (or dynamic instrumentation)}~\cite{valgrind, pin, DynamoRIO, perkins2009automatically} modifies the binary's behaviors 
during its execution through the injection of instrumentation code.
Due to the need to instrument the code at runtime, this technique may result in higher execution time compared to the original binary. 
The main advantage of dynamic rewriting is its ability to accurately capture information about a program's states or behaviors,
which is much harder when using  static rewriting.
Example dynamic binary rewriting tools include \texttt{Pin}~\cite{pin} and \texttt{DynamoRIO}~\cite{DynamoRIO}.

In this work, we first leverage the runtime information retrieved from dynamic instrumentation to accurately locate instructions that need to be rewritten.
Instruction rewriting is then performed statically in order to minimize the runtime overhead of the rewritten binary.

\ignore{
\begin{figure}[t]
	\centering
	\includegraphics[width=.7\linewidth]{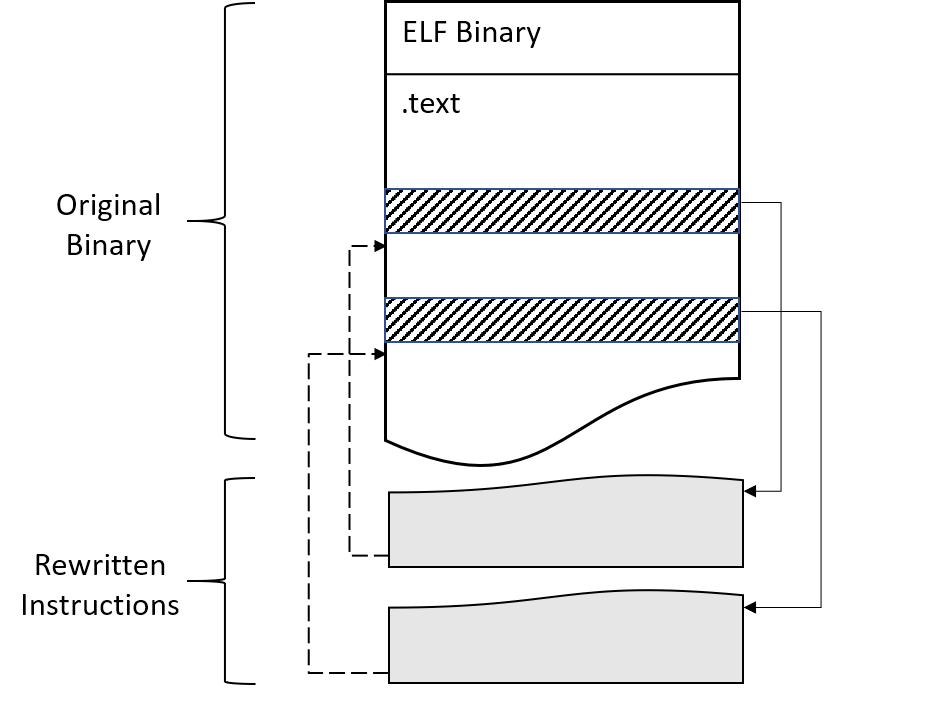}
	\caption{Illustration of static binary rewriting strategy. 
	An arrow represent an unconditional jump.}
	\label{fig:static-rewriting}
\end{figure}
}


\section{Overview of the \acron Framework\label{sec:approach-overview}}

The most straightforward, and obvious, approach to replace implementations of vulnerable cryptographic primitives 
requires modifying (and then recompiling) a program's source code. This takes 
time and effort, and renders it difficult to fix legacy software for 
which source code may not be available. Instead, we propose \acron\ -- a toolchain that automatically augments 
and replaces weak, vulnerable, and/or broken cryptographic primitives at the binary level.  

To better illustrate how \acron works, we start by presenting a simple representative example shown in Figure~\ref{fig:crypto-example}.
This example program first computes an MD5 digest over an input string.
The digest is then converted into a human-readable form, which is in turn displayed to the user.

MD5 has been shown to be vulnerable to collision and pre-image attacks~\cite{klima2006tunnels, sasaki2009finding}.
Suppose that a system or software developer would like to manually rewrite parts of the binary
in order to support a more secure hash algorithm -- e.g., SHA-256.
One way to accomplish this task is to perform the following steps:

\noindent {\bf Step-1:} Identify the functions in the binary that implement MD5.

\noindent {\bf Step-2:}  Recover the type and order of parameters in the identified functions.

\noindent {\bf Step-3:} Insert an implementation of a SHA-256 function with the same type and order of parameters into the original binary.

\noindent {\bf Step-4:} Redirect all calls to MD5 to the newly added SHA-256 function.

\noindent {\bf Step-5:}  Determine all changes throughout the binary affected by an increase in the digest size (MD5's digest size is 128 bits while that of SHA-256 is 256 bits).

\noindent {\bf Step-6:}  Rewrite the binary according to changes discovered in step-5.

\begin{figure*}[!h]
\begin{minipage}{.48\linewidth}
\begin{lstlisting}[style=CStyle]
void MD5(const unsigned char* input, size_t inputlen, unsigned char* output) {
	MD5_CTX ctx;
	MD5Init(&ctx);
	MD5Update(&ctx, input, inputlen);
	MD5Final(output, &ctx);
}

int main(void) {
	// Initialize input and output buffers
	char input[] = "Hello, world!";
	size_t inputlen = strlen(input);
	unsigned char digest[16];
	char hexdigest[33] = {0};
	
	// Compute: output = MD5(input)
	MD5(input, inputlen, digest);
	
	// Convert output digest to hex string
	for(int i=0; i < 16; i++) {
		sprintf(hexdigest+2*i, "%02x", digest[i]);
	}
	
	// Print digest in hex format
	for(int i=0; i < 16; i++) {
		printf("%02x", digest[i]);
	}
	
	// Print hexdigest string
	printf("\n%s\n", hexdigest);
	return 0;
}
\end{lstlisting}
\centering
{\scriptsize (a) Simple program utilizing a cryptographic primitive (the MD5 hash function)}
\end{minipage}
\hfill
\begin{minipage}{.48\linewidth}
\lstset{language=[x64]Assembler,
	frame=single, 
        basicstyle=\tiny\ttfamily, 
        keywordstyle=\color{Blue}, 
	 morecomment=[l][\color{Orange}]{;},
	 escapechar=\&,
        tabsize=5, 
        numbers=none, 
        firstnumber=1, 
        stepnumber=5 
        }
\begin{lstlisting}
<main_fn_prologue>:
  400629:  push   rbp
  40062a:  mov    rbp,rsp
  40062d:  sub    rsp,0x60					
  ...
<call_to_md5>:
  400684:  call   4004c0 <strlen@plt>
  400689:  mov    rcx,rax
  40068c:  lea    rdx,[rbp-0x40]				
  400690:  lea    rax,[rbp-0x50]				
  400694:  mov    rsi,rcx
  400697:  mov    rdi,rax
  40069a:  call   400616 <MD5>
  ...
<sprintf_loop_body>:
  4006ad:  movzx  eax,BYTE PTR [rbp+rax*1-0x40]	
  4006b2:  movzx  eax,al
  4006b5:  mov    edx,DWORD PTR [rbp-0x54]		
  4006b8:  add    edx,edx
  4006ba:  movsxd rdx,edx
  4006bd:  lea    rcx,[rbp-0x30]				
  4006c1:  add    rcx,rdx					
  4006c4:  mov    edx,eax
  4006c6:  mov    esi,0x4007d4
  4006cb:  mov    rdi,rcx
  4006ce:  mov    eax,0x0
  4006d3:  call   400500 <sprintf@plt>
<sprintf_loop_condition>:
  4006d8:  add    DWORD PTR [rbp-0x54],0x1		
  4006dc:  cmp    DWORD PTR [rbp-0x54],0xf
  4006e0:  jle    4006a8 <main+0x7f>
  ...
<printf_loop_body>:
  4006eb:  mov    eax,DWORD PTR [rbp-0x54]
  4006ee:  cdqe   
  4006f0:  movzx  eax,BYTE PTR [rbp+rax*1-0x40]	
  4006f5:  movzx  eax,al
  4006f8:  mov    esi,eax
  4006fa:  mov    edi,0x4007d4
  4006ff:  mov    eax,0x0
  400704:  call   4004e0 <printf@plt>
<printf_loop_condition>:
  400709:  add    DWORD PTR [rbp-0x54],0x1		
  40070d:  cmp    DWORD PTR [rbp-0x54],0xf
  400711:  jle    4006eb <main+0xc2>
\end{lstlisting}
	\centering

	{\scriptsize (b) Corresponding disassembly of \texttt{main} function in (a), compiled with \texttt{O0} flag}
\end{minipage}

\caption{An example of a simple program utilizing a cryptographic primitive (the MD5 hash function in this case)}
\label{fig:crypto-example}
\vspace{-.5cm}
\end{figure*}

~\\
\noindent\textbf{Goal \& Scope:} 
\acron is designed to automate the aforementioned steps. 
It targets ELF-based X86/64 binaries generated by compiling \texttt{C} programs.
We do not assume any knowledge of, or require, the corresponding source code or debugging symbols.
Since \acron is built as a defensive tool to work on standard and legacy software,
it assumes the target programs are not malicious. \emph{Obfuscated or malware binaries are out of scope in this work.}
We demonstrate concrete feasibility 
on cryptographic hash functions,
but the design and ideas behind \projectname are general and can be 
applied to other primitives too.

\noindent\underline{NOTE:} We use \textit{cryptographic hash functions} (or \textit{hash functions})
to denote the algorithmic details behind the implementations/executables.
We denote the functions (or methods) in such implementations/executables that realize 
such hash function(s) as \textit{hash routines}.
We use \textit{target} hash functions/routines to refer to
the insecure hash functions/routines that need to be identified and replaced.


\section{Design Details of \acron \label{sec:alice-details}}

This section discusses the design details of the \acron toolchain.
The operation of \acron consists of three main phases:
(i) identifying cryptographic primitives, (ii) scoping changes, and (iii) augmenting and rewriting changes.

\subsection{Identifying Cryptographic Primitives (Hash Functions)\label{sec:ident}}

We designed \acron to target non-malicious (i.e., unobfuscated) binary programs.
The first phase of \acron leverages this characteristic and identifies hash functions by first detecting static features that are known of the target hash function.

\begin{observation}[Constants]\label{obv:const}
	A common design approach for hash functions is to 
		initialize a digest buffer using well-known constants. As an example, MD5 uses 32-bit constant words: $X_1 | X_2 | X_3 | X_4$ 
		with the following hex values: \\
		\begin{center}
		\vspace{-.2cm}
		$X_1 = \texttt{0x67452301}$, $X_2 = \texttt{0xEFCDAB89}$, $X_3 = \texttt{0x98BADCFE}$, $X_4 = \texttt{0x10325476}$
		\end{center}
\end{observation}

If we locate such constants in a binary,
there is a high chance that a routine enclosing those constants implements part(s) of the MD5 hash function.
Our approach starts by scanning a binary program to find the addresses where known constants appear.
We then mark a routine in which those constants are enclosed as a candidate implementation of the target hash function.

\begin{observation}[Context Initialization]
	The example in Figure~\ref{fig:crypto-example}a illustrates a typical usage of 
		a hash function in practice. An application function -- \texttt{main()} -- calls the \texttt{MD5()} function, which in turn
		calls \texttt{MD5Init()} to initialize a digest buffer with known constants.
		Having a dedicated function to setup an initial context (e.g., \texttt{MD5Init()}) is common practice when implementing most cryptographic primitives
		and can be found in several open-source libraries such as the OpenSSL or libgcrypt.
\end{observation}

This observation suggests that the identified candidate implementation will typically correspond to the initialization routine -- \texttt{Init()}.
However, it is not always the case as \texttt{Init()} could be inlined.
For example, \texttt{MD5Init()} in Figure~\ref{fig:crypto-example} will be inlined inside \texttt{MD5()} 
when the program is compiled with the optimization flag O3.
In this scenario, the identified routine will instead correspond to the implementation of the target hash function -- \texttt{MD5()}.
One could use a simple heuristic based on the size of the routine to distinguish between the two scenarios. 
However, we found this approach to produce a lot of false-negatives in practice.
Instead, we adopt a more conservative approach and consider routines produced in both scenarios as candidates.
More specifically, \acron analyzes the program's callgraph and determines all routines that invoke the previously identified routine.
It then includes those caller routines into the list of candidates.

Now, \acron needs a mechanism to eliminate false-positives,
which can arise due to two reasons.
The first reason is that our approach so far focuses on ensuring no false-negatives by accepting all routines possibly implementing a target hash routine.
The second reason is that static features such as a constant vector are not always unique to a single hash function.
It is not uncommon for different hash functions to share the same constant vectors.
Examples of a pair of hash functions that use the same constant vectors are
BLAKE2b -- SHA-512 and MD5 -- MD4.
In Figure~\ref{fig:crypto-example}, even if we successfully determine that \texttt{MD5Init()} is inlined,
we still cannot easily distinguish whether the identified hash routine implements MD4 or MD5 hash function.

\begin{observation}[Input/Output Uniqueness]
A cryptographic hash function is deterministic, i.e., for a given input string, it always generates the same digest as output.
The input/output pair is usually (in practice) unique to the hash function that produces them.
\end{observation}

With this observation, the best way to test whether
a candidate implements the target hash function is to execute the identified routine,
and compare the resulting output with the expected output.
Since we expect the identification phase to be an offline computation,
naturally we would base this step of our approach on an offline dynamic execution technique,
which allows us to execute a given routine with any concrete chosen input. 
This is in contrast with online dynamic execution, which requires running the entire binary program with test cases.
To perform offline dynamic execution,
it is necessary to setup a call stack with proper parameters that will be passed into that routine.

\begin{observation}[Parameters]
An implementation of a cryptographic function generally takes a fixed number of function parameters.
For example, in Figure~\ref{fig:crypto-example}a, \texttt{MD5()} include three parameters: an input string, the input length and an output digest.
Some implementations do not mandate the use of input length as it can be inferred from the input string (e.g., via \texttt{strlen()}).
Thus, a hash routine in such implementations will take only two parameters.
It is also worth noting that, even though a number is fixed, the order in which these parameters appear may not be the same for all implementations.
\end{observation}

Based on this observation, \acron enumerates all possible combinations of a hash routine's parameters and
prepares 8 ( = 3!+2!) call stacks, each initialized with different combinations of parameters.
It then executes a candidate routine on each call stack and observes the output buffer after the execution.
The first phase of \acron finishes by outputting candidates producing the expected output
as well as the parameter information obtained from corresponding call stacks.


\subsection{Scoping Changes}\label{sec:scope}
After locating  target hash routines, \acron must determine changes (throughout the binary) that are required for replacing such routines.
We now describe three categories of such changes  
using the illustrative example in Figure~\ref{fig:crypto-example},
and later outline how to identify a subset of such changes (at the binary level). 

\begin{itemize}
	\item[\fbox{\bf C1}] {\bf Routine Replacement:} The first category is a change in the hash routine itself.
	 Code/instructions  implementing the target hash routine are replaced by code/instructions that implement a more secure hash function.
	For example, if our goal is to replace the \texttt{MD5} function with \texttt{SHA-256} in Figure~\ref{fig:crypto-example}, 
	instructions corresponding to \texttt{MD5()} (i.e., the ones starting from address \texttt{0x400616})
	need to be replaced by \texttt{SHA-256} instructions.
	We will discuss how \acron augments this type of change into a binary in Section~\ref{sec:rewrite}.
	
	\item[\fbox{\bf C2}] {\bf Changes in Buffers Sizes:}
	Depending on the digest size of both the replacement and the target hash functions, 
	other related memory buffers may need to be enlarged to correctly accommodate the new replacement routine.
	For instance, replacing \texttt{MD5} with \texttt{SHA-256} in Figure~\ref{fig:crypto-example} 
	would also require enlarging the
	size of variables storing the output of the hash function (i.e., \texttt{digest} variable) from 16 bytes to 32 bytes.  
	This change in buffer size affects other memory buffers
	that consume the output digest, e.g., \texttt{hexdigest} also needs to be expanded by 16 bytes. 
	Such changes have to be scoped and propagated throughout the entire binary.
	We discuss how to identify this type of change in the remaining of this section and how \acron performs
	augmentation on such changes in Section~\ref{sec:rewrite}.
	
	\item[\fbox{\bf C3}] {\bf Changes in Logic:} 
	This category refers to changes that need to be applied to the underlying binary logic in order to have
	a correct resulting binary function.
	For example, in Figure~\ref{fig:crypto-example}, simply replacing \texttt{MD5()} with \texttt{SHA-256()} and enlarging \texttt{hexdigest} and \texttt{digest} variable
	do not suffice to produce the desired binary.
	One would have to also edit a loop terminating condition from $i < 16$ to $i < 32$ in line 19 and 24 to reflect the replacement \texttt{SHA-256}.
	At the binary-level, this change corresponds to modifying the instructions at addresses \texttt{0x4006dc} and \texttt{0x40070d} from [\texttt{cmp    DWORD PTR [rbp-0x54],0xf}]
	to [\texttt{cmp    DWORD PTR [rbp-0x54],0x1f}]. 
	This requires knowing that the constant \texttt{0xf} in those instructions is related to the digest size.
	However, it is hard, in some cases impossible, to locate and augment this type of changes without any prior knowledge of the correct behavior of the resulting binary.
	Therefore, we do not consider this category of changes in this work. 
	We further discuss the difficulty of automatically determining this type of change in Appendix~\ref{apdx:limitation}.
\end{itemize}

Of the three categories of changes, \fbox{\bf C1} is identified in the previous phase of \acron in Section~\ref{sec:ident}
while \fbox{\bf C3} is out of scope in this work.
The remainder of this section will focus on how \acron locates the changes from \fbox{\bf C2}.

\acron leverages dynamic taint analysis to determine the change in buffer size \fbox{\bf C2}.
Typically, dynamic taint analysis starts by marking any data that comes from an untrusted source as tainted. 
It then observes program execution to keep track of the flow of tainted data in registers and memory.
We adapt this idea to identify all \textit{memory buffers} that are affected (or tainted) by the output digest of the target hash routine. 
In particular, our dynamic taint analysis executes a binary on test inputs with the following taint policies:\\

\noindent\textbf{Taint introduction.}
At the beginning of execution, \acron initializes all memory locations to be non-tainted.
During the execution, whenever entering the target hash routine, 
\acron reads the value in the parameter registers to observe the base address of the output digest.
Since the digest size is well-known and deterministic for any given hash function,
\acron can also identify the entire address range of the digest buffer.
Upon exiting the routine, \acron then assigns a taint label to all memory locations in the digest buffer. 
\acron uses three taint labels to differentiate 3 types of memory allocations:

\begin{itemize}
	\item \textbf{Static allocation.} In our target executables, static memory is allocated at compile time
	before the program is executed. Thus, the location of this type of memory is usually deterministic, stored
	in either \texttt{.data} or \texttt{.bss} segment of the associated binary.
	Detecting whether a given memory location is statically allocated is simply done by checking whether its address
	lies within the boundaries of those segments.
	
	\item \textbf{Heap-based allocation.}
	\acron traces all heap-based memory allocations by intercepting a call to three well-known 
	\texttt{C} routines: \texttt{malloc()}, \texttt{calloc()} and \texttt{realloc()}.
	Whenever each of these routines is called, \acron learns the size of allocated memory by reading values of its parameter registers.
	Upon exiting the same routine, \acron then learns the base address of allocated memory via the return value. 
	With this information, \acron later can determine whether memory at a given location is allocated on the heap.
	
	\item \textbf{Stack-based allocation.} 
	\acron maintains stack-related information of the execution via a \textit{shadow} stack. 
	Specifically, after executing any \texttt{call} instruction, \acron pushes into the shadow stack: a pair of the current stack pointer and an address of the function being called. 
	Upon returning from a routine (via a \texttt{ret} instruction), \acron pops the shadow stack.
	This information allows \acron to reconstruct stack frames at any point during the execution of dynamic taint analysis.
	\acron determines whether memory at a given address is on the stack
	by checking it against all stack frames. 
\end{itemize}

\noindent\textbf{Taint Propagation.}
\acron's taint propagation rules are enforced at the word-level granularity.
While we could use a more precise granularity such as the bit-level~\cite{yadegari2014bit},
we did not find such approach to be cost-effective
as test inputs may require a timely interaction with a remote server;
having a significantly long delay in the dynamic taint analysis can cause the remote server to timeout 
and consequently the analysis may not be performed as expected.

In addition to the general taint propagation rules, 
\acron also considers the \textit{taint-through-pointer} scenario:
if a register \texttt{A} is tainted and a register \texttt{B}
is assigned with the referenced value of \texttt{A}, i.e., \texttt{B := *A}, 
then \texttt{B} is considered tainted.
Such rule is necessary to accurately capture the data-flow in a common usage of a hash function,
where the raw digest value is converted to human-readable format via a look-up table, e.g.,
the use of \texttt{sprintf()} in line 20 of Figure~\ref{fig:crypto-example}a.

Using these rules, \acron's dynamic taint analysis
can determine, and assign taint labels to, all memory locations affected by the output digest.
At the end of the analysis,
\acron aggregates individual tainted memory locations into unified memory buffers.
Our aggregation rule is simple: \acron considers contiguous memory locations to be 
a memory buffer if their address range is at least as long as the target hash function's digest size.
Lastly, in this phase, \acron outputs the types (i.e., either stack-based, heap-based or static), locations (e.g., a stack offset or global address), 
and relevant instructions (e.g., an instruction address of a call to \texttt{malloc}) of memory buffers that are derived from the output digest. 


\subsection{Augmenting and Rewriting Changes}\label{sec:rewrite}


\acron can incorporate several  rewriting approaches.
Since runtime of rewritten binaries is our primary concern, 
we mainly use static binary rewriting that has been previously shown to have 
minimal impact on the runtime~\cite{bauman12superset}.

To reduce the size of rewritten binaries,
we rewrite at \textit{routine level} rather than at section level\footnote{We intentionally avoid rewriting at the instruction level as this can potentially incur significant run-time overhead for the rewritten/output binaries.}.
If there is at least one instruction that needs to be edited in a particular routine,
we rewrite the binary as follows: 
\begin{enumerate}
	\item Create a new empty section in the binary
	\item Apply changes from \fbox{\bf C1} and \fbox{\bf C2} to the routine
	\item Modify the routine with respect to the placement of the new section
	\item Insert the entire rewritten routine into the new section
	\item Insert a \texttt{jump} instruction to the new section at original routine's entry point 
\end{enumerate}
Steps (1), (4) and (5) are explained in Section~\ref{sec:rel-work}. We focus in this section on steps (2) and (3).
We refer to Appendix~\ref{sec:impl} for implementation details of all steps.

For step (3), we only need to ensure that the rewritten routine maintains the correct control flow targets.
Doing so requires editing all instruction operands in the routine that use \texttt{rip}-relative addressing.
The displacement of such operands is recomputed based on the address of the new location:
\begin{center}
$new\_disp = old\_disp + old\_inst\_addr - new\_inst\_addr$
\end{center}

In step (2), to apply changes from \fbox{\bf C1}, \acron generates a patch from a user-supplied \texttt{C} code that implements the replacement hash function,
and adds them to the new empty section of the binary.
To ensure correctness of the rewritten binaries, implementation of the user-supplied replacement hash function must have the same parameter order as that of the target hash function as well as be self-contained.
These requirements are, however, simple to fulfill given that the user already has access to the replacement hash function's source-code; we discuss this issue in detail in Appendix~\ref{sec:impl}.
We also ensure that a call to the target hash routine is redirected to this new code
by simply rewriting the first instruction of the target hash routine to: \texttt{jmp [new\_code\_entry\_point]}.
For each memory buffer identified in \fbox{\bf C2}, \acron 
computes the new buffer size based on the ratio of the digest sizes of the target hash function and that of the replacement hash function, i.e., $new\_size = \lceil{}old\_size \times |digest_{secure}|/|digest_{target}| \rceil{}$.
\acron rewrites the binary to support the expanded buffers by
employing different techniques for each type of buffers:

\noindent{\bf Static Buffer.} As a static buffer is allocated at a fixed address,
	we expand such buffer by creating another buffer at a new location and modify all instruction operands that access the original buffer to this newly allocated buffer.
	Specifically, \acron first allocates a new data segment in the binary, and
	creates a mapping of the address of the original buffer to the address in the new data segment.
	To ensure that the rewritten binary uses the new address instead of the original, \acron scans through all instructions in the original binary
	and edits the ones that contain an access to the original address by using information obtained from the previously computed address mapping.
	
\noindent{\bf Heap-based buffer.} Unlike a static buffer, this type of buffer is allocated dynamically through a call to \texttt{malloc()}, \texttt{alloc()} or  \texttt{realloc()} routine.
	Fortunately, \acron learns when this type of buffer is allocated through the dynamic taint analysis in Section~\ref{sec:scope}.
	Thus, expanding a heap-based buffer only requires \acron to trace back to the instruction allocating such buffer, i.e. a \texttt{call} instruction to \texttt{malloc()}, \texttt{alloc()} or  \texttt{realloc()},
	and update the parameter register value storing the allocation size information to the new buffer size.

\begin{figure*}[!hbt]
	\centering
	\begin{minipage}{.48\linewidth}
\lstset{language=[x64]Assembler,
	frame=single, 
        basicstyle=\tiny\ttfamily, 
        keywordstyle=\color{Blue}, 
	 morecomment=[l][\color{Orange}]{;},
	 escapechar=\&,
        tabsize=5, 
        numbers=none, 
        firstnumber=1, 
        xleftmargin=0.1\linewidth,
        xrightmargin=0.1\linewidth
        }
\begin{lstlisting}[linebackgroundcolor={\lstcolorlines[lgreen]{4,9,10,16,18,21,26,27,31,33,40,41}}]
<main_fn_prologue>:
  400629:  push   rbp
  40062a:  mov    rbp,rsp
  40062d:  sub    rsp,0x60					
  ...
<call_to_md5>:
  400684:  call   4004c0 <strlen@plt>
  400689:  mov    rcx,rax
  40068c:  lea    rdx,[rbp-0x40]				
  400690:  lea    rax,[rbp-0x50]				
  400694:  mov    rsi,rcx
  400697:  mov    rdi,rax
  40069a:  call   400616 <MD5>
  ...
<sprintf_loop_body>:
  4006ad:  movzx  eax,PTR [rbp+rax*1-0x40]	
  4006b2:  movzx  eax,al
  4006b5:  mov    edx,DWORD PTR [rbp-0x54]		
  4006b8:  add    edx,edx
  4006ba:  movsxd rdx,edx
  4006bd:  lea    rcx,[rbp-0x30]				
  4006c1:  add    rcx,rdx					
  ...
  4006d3:  call   400500 <sprintf@plt>
<sprintf_loop_condition>:
  4006d8:  add    DWORD PTR [rbp-0x54],0x1		
  4006dc:  cmp    DWORD PTR [rbp-0x54],0xf
  4006e0:  jle    4006a8 <main+0x7f>
  ...
<printf_loop_body>:
  4006eb:  mov    eax,DWORD PTR [rbp-0x54]
  4006ee:  cdqe   
  4006f0:  movzx  eax,PTR [rbp+rax*1-0x40]	
  4006f5:  movzx  eax,al
  4006f8:  mov    esi,eax
  4006fa:  mov    edi,0x4007d4
  4006ff:  mov    eax,0x0
  400704:  call   4004e0 <printf@plt>
<printf_loop_condition>:
  400709:  add    DWORD PTR [rbp-0x54],0x1		
  40070d:  cmp    DWORD PTR [rbp-0x54],0xf
  400711:  jle    4006eb <main+0xc2>
\end{lstlisting}

		\centering
		(a) Original Binary
	\end{minipage}
	\hfill
	\begin{minipage}{.48\linewidth}
	
\lstset{language=[x64]Assembler,
	frame=single, 
        basicstyle=\tiny\ttfamily, 
        keywordstyle=\color{Blue}, 
        morekeywords=[2]{x50,x60,x64,x70,x80,x84,x90},
        keywordstyle=[2]\color{Red}\bf, 
	 morecomment=[l][\color{Orange}]{;},
        tabsize=5, 
        numbers=none, 
        firstnumber=1, 
        xleftmargin=0.1\linewidth,
        xrightmargin=0.1\linewidth
        }
\begin{lstlisting}[linebackgroundcolor={\lstcolorlines[lgreen]{4,9,10,16,18,21,26,27,31,33,40,41}}]
<main_fn_prologue>:
  400629:  push   rbp
  40062a:  mov    rbp,rsp
  40062d:  sub    rsp,0x90					
  ...
<call_to_md5>:
  400684:  call   4004c0 <strlen@plt>
  400689:  mov    rcx,rax
  40068c:  lea    rdx,[rbp-0x70]				
  400690:  lea    rax,[rbp-0x80]				
  400694:  mov    rsi,rcx
  400697:  mov    rdi,rax
  40069a:  call   400616 <MD5>
  ...
<sprintf_loop_body>:
  4006ad:  movzx  eax,PTR [rbp+rax*1-0x70]	
  4006b2:  movzx  eax,al
  4006b5:  mov    edx,DWORD PTR [rbp-0x84]		
  4006b8:  add    edx,edx
  4006ba:  movsxd rdx,edx
  4006bd:  lea    rcx,[rbp-0x50]				
  4006c1:  add    rcx,rdx					
  ...
  4006d3:  call   400500 <sprintf@plt>
<sprintf_loop_condition>:
  4006d8:  add    DWORD PTR [rbp-0x84],0x1		
  4006dc:  cmp    DWORD PTR [rbp-0x84],0xf
  4006e0:  jle    4006a8 <main+0x7f>
  ...
<printf_loop_body>:
  4006eb:  mov    eax,DWORD PTR [rbp-0x84]
  4006ee:  cdqe   
  4006f0:  movzx  eax,PTR [rbp+rax*1-0x70]	
  4006f5:  movzx  eax,al
  4006f8:  mov    esi,eax
  4006fa:  mov    edi,0x4007d4
  4006ff:  mov    eax,0x0
  400704:  call   4004e0 <printf@plt>
<printf_loop_condition>:
  400709:  add    DWORD PTR [rbp-0x84],0x1		
  40070d:  cmp    DWORD PTR [rbp-0x84],0xf
  400711:  jle    4006eb <main+0xc2>
\end{lstlisting}
		\centering
		(b) Rewritten Binary
	\end{minipage}
	\caption{Disassembly of \texttt{main} before \& after increasing \texttt{digest} and \texttt{hexdigest} buffers by 16 and 32 bytes, respectively. 
	Lines containing rewritten instructions are highlighted in \textcolor{green}{green} and changes are in \textcolor{red}{red}. 
	}
	\label{fig:stack-change}
	\vspace{-.5cm}
\end{figure*}

\noindent{\bf Stack-based buffer.} 
Figure~\ref{fig:stack-change} shows how \acron modifies the \texttt{main} routine in the example from Figure~\ref{fig:crypto-example}b
to support the expansion of stack-based buffers: \texttt{digest} and \texttt{hexdigest} by 16 and 32 bytes respectively.
Intuitively, expanding a buffer allocated on the stack at the binary level requires:
(i) locating the routine that uses the corresponding stack frame, (ii) enlarging the frame to be large enough to hold the new buffers, and
(iii) adjusting every access to memory inside the frame accordingly.
\acron's previous phase, in Section~\ref{sec:scope}, provides necessary information to satisfy the first requirement (via the shadow stack).
To achieve the second requirement, \acron rewrites the instructions that are responsible
for increasing and decreasing the stack pointer in the prologue/epilogue of the located routine, e.g., \texttt{[40062d: sub rsp,0x60]} in Figure~\ref{fig:stack-change}a. 
For the third requirement, \acron iterates through all instructions in the routine and
inspects the ones that use the stack offset, i.e., via \texttt{rsp} or \texttt{rbp} registers.
\acron then recomputes the stack offset with respect to the increased frame size
and rewrites those instructions if the newly computed offset differs from the original.
Steps of how \acron recomputes the new stack offset are shown in Algorithm~\ref{alg:stack-offset} of Appendix~\ref{apdx:alg}.
In Figure~\ref{fig:stack-change}, \acron identifies all instructions that access a stack element and rewrites the ones highlighted in green. 

\ignore{
\begin{remarkk}
Out of three types, we expect to encounter the stack-based buffer the most
while the heap-based buffer is less likely.
This is because developers likely declare a digest buffer (or any buffer tainted by this buffer) as a fixed-length array
due to its deterministic size, which is naturally unsuitable for heap-based allocation.
This is evident from the datasets used for our evaluation in Section~\ref{sec:eval}.
We thus exclude the implementation of the rewriting rules for the heap-based buffer in \acron and only focus
on the other two types.
\end{remarkk}
For each instruction in replaced function, we rewrite:
\begin{itemize}
	\item if function needs to change stack size and instruction is in prologue and epilogue expanding stack size, change to new size. Delta size = sum of delta buffer size for all buffers that need to be expanded in that function.
	\item if it accesses stack, re-compute its displacement, modify if necessary.
	\item if it accesses tainted static memory, use the new address
	\item if refer relative address (i.e., rpi + x), recompute the relative address
	\item if it is a jump and jump target points to replaced function, recompute jump target.
\end{itemize}
}


\section{Experimental Evaluation \label{sec:eval}}

\subsection{Experimental Setup}

\noindent\textbf{Goals \& Datasets.} 
The goal of this evaluation is two-fold, first, to assess whether \acron can accurately identify and replace different implementations of hash functions, and second, to measure \acron's effectiveness on real-world applications.
Different implementations may include different hash function structures (e.g., with or without \texttt{Init()}), different parameter orders, or simply different implementation details.
We apply \acron to a dataset that consists of four popular cryptographic libraries: \texttt{OpenSSL}, \texttt{libgcrypt}, \texttt{mbedTLS},
and \texttt{FreeBL}.
We compile each library with different optimization levels, including \texttt{O0}, \texttt{O1}, \texttt{O2}, \texttt{O3} and \texttt{Os}, into a static library.
We then create a simple \texttt{C} application (similar to the one in Figure~\ref{fig:crypto-example}a) that calls exactly one hash function located in the static library.
We compile this application without debugging/relocation symbols and link it with each individual static library.
\karim{We also assess \acron's effectiveness on 6 real-world applications: \texttt{smd5\_mkpass} and \texttt{ssha\_mkpass} -- github projects for creating LDAP passwords~\footnote{https://github.com/pellucida/ldap-passwords}, \texttt{md5sum} and \texttt{sha1sum} -- string/file checksum programs, \texttt{lighttpd} -- a lightweight webserver program, and \texttt{curl} -- a webclient command line tool.}
Similar to the first dataset, each program was compiled without debugging symbols and with various optimization levels,
and statically linked to the cryptographic library used within the program. 

\noindent\textbf{Insecure Hash Functions.} We consider MD2, MD4, MD5, SHA1, and RIPEMD-160 as insecure hash functions to be replaced in our experiments. Our objective is to identify implementations of such hash functions in binaries and replace them with stronger ones, i.e., SHA-256.
A list of all (insecure) hash functions in each dataset is shown in Table~\ref{tab:cryptolib}.

\noindent\textbf{Environment.} Experiments are performed on a virtual machine with Ubuntu 16.04.5 OS, 4GB of RAM and 2 cores of 3.4GHz of CPU
running on top of an Intel i7-3770 machine. The following versions of required tools are used in the experiments:
\texttt{gcc-5.4.0}, \texttt{angr-7.8.2.21}~\cite{angr}, \texttt{Triton-0.6}~\cite{triton} and \texttt{Pin-2.14.71313}~\cite{pin}.

\begin{table}[]
\centering
\resizebox{.8\columnwidth}{!}{%
\begin{tabular}{l|c|ccccc}
      \textbf{Dataset}  &  \textbf{Version} & \textbf{MD2}                   & \textbf{MD4}                   & \textbf{MD5}                   & \textbf{SHA1}                  & \textbf{RIPEMD160}             \\ \hline
\multirow{4}{*}{Crypto Libraries} & OpenSSL-1.1.1   & \xmark & \texttt{ILO}  & \texttt{ILO}  & \texttt{ILO}  & \texttt{ILO}  \\
& libgcrypt-1.8.4 & \xmark & \xmark & \xmark & \texttt{OIL} &  \texttt{OIL} \\
& mbedTLS-2.16.0   & \texttt{ILO} & \texttt{ILO} & \texttt{ILO} & \texttt{ILO} & \xmark \\
& FreeBL-3.42    &  \texttt{OI} & \xmark &  \texttt{OIL},\texttt{OI} &  \texttt{OIL},\texttt{OI} & \xmark \\ \cdashline{1-7}
\multirow{6}{*}{Real-world Programs} 
& \texttt{smd5\_mkpass} & \xmark & \xmark & \texttt{OIL} & \xmark & \xmark \\
& \texttt{ssha\_mkpass} & \xmark & \xmark & \xmark & \texttt{OIL} & \xmark \\
& \texttt{md5sum-5.2.1} & \xmark & \xmark & \texttt{ILO} & \xmark & \xmark \\
& \texttt{sha1sum-5.2.1} & \xmark & \xmark & \xmark & \texttt{ILO} & \xmark \\
& \texttt{curl-7.56.0} & \xmark & \xmark & \texttt{OI} & \xmark & \xmark \\
& \texttt{lighttpd-1.4.49} & \xmark & \xmark & \xmark & \texttt{ILO} & \xmark
\end{tabular}
}
\caption{Hash functions used in our test datasets.  
		\xmark\ indicates no hash function while \texttt{ILO}, \texttt{OIL} and \texttt{OI} denote a function with the parameter 
		order \texttt{(input,inputlen,output)}, \texttt{(output,input,inputlen)} and \texttt{(output,input)}. 
}
\label{tab:cryptolib}
\vspace{-1cm}
\end{table}

\subsection{Evaluation Results: Cryptographic Libraries}\label{sec:eval_lib}

As described in Section~\ref{sec:scope}, we only consider automated inference of required changes from categories \fbox{\bf C1} and \fbox{\bf C2} in this work.
In order to properly evaluate \acron, we perform manual analysis 
to identify all changes required in \fbox{\bf C3} and supply them to \acron.
The manually supplied changes for this case (in this dataset) consist of only a couple of instructions that typically specify loop termination condition(s).
For example, in the binary from Figure~\ref{fig:crypto-example}b, we instruct \acron 
to modify two instructions at addresses \texttt{0x4006dc} and \texttt{0x40070d}
from \texttt{[cmp    DWORD PTR [rbp-0x54],0xf]} 
to \texttt{[cmp    DWORD PTR [rbp-0x54],0x1f]}.

\noindent\textbf{Correctness of Rewritten Binary.} 
To simplify illustration, we first describe behaviors of the rewritten binaries that are considered  \emph{incorrect} in this dataset.
First, if \acron misidentifies any of necessary changes in the input binary, the resulting binary will not display
the \emph{correct} SHA-256 digest of the \texttt{input} variable. For instance, it may display nonsensical data, a digest produced by the original hash function
or an incomplete version of the SHA-256 digest.
Second, if \acron's rewriting phase does not function properly (e.g., it expands the buffer size by a different amount or adjusts memory access to the stack incorrectly),
it likely results in a runtime error for the output binary.
We consider the correctness of the rewritten binaries from this dataset to be the converse of the aforementioned behaviors, i.e., 
execution of the output binary must terminate without any errors and it must result in displaying the correct SHA-256 digest of the \texttt{input} variable.
All binaries produced by \acron in this dataset work as expected.

\noindent\textbf{Binary Size and Execution Overhead.}
\acron adds around 3-13KB to the output binaries (see breakdown details in Appendix~\ref{apdx:size_overhead}).
On further inspection, we found two main reasons for this overhead.
First, \acron statically adds a patch implementing the replacement SHA-256 hash function, which contributes around 3KB to the output binary.
Second, our underlying binary rewriter, \texttt{patchkit}~\cite{patchkit}, expects code and data of this patch to be aligned to a page size (i.e., 4KB in our testing machine),
which can add up to another 8KB to the output binary.

\begin{figure}[!h]
	\centering
	\subfloat[Number of executed instructions.]{
  		\includegraphics[width=.5\linewidth]{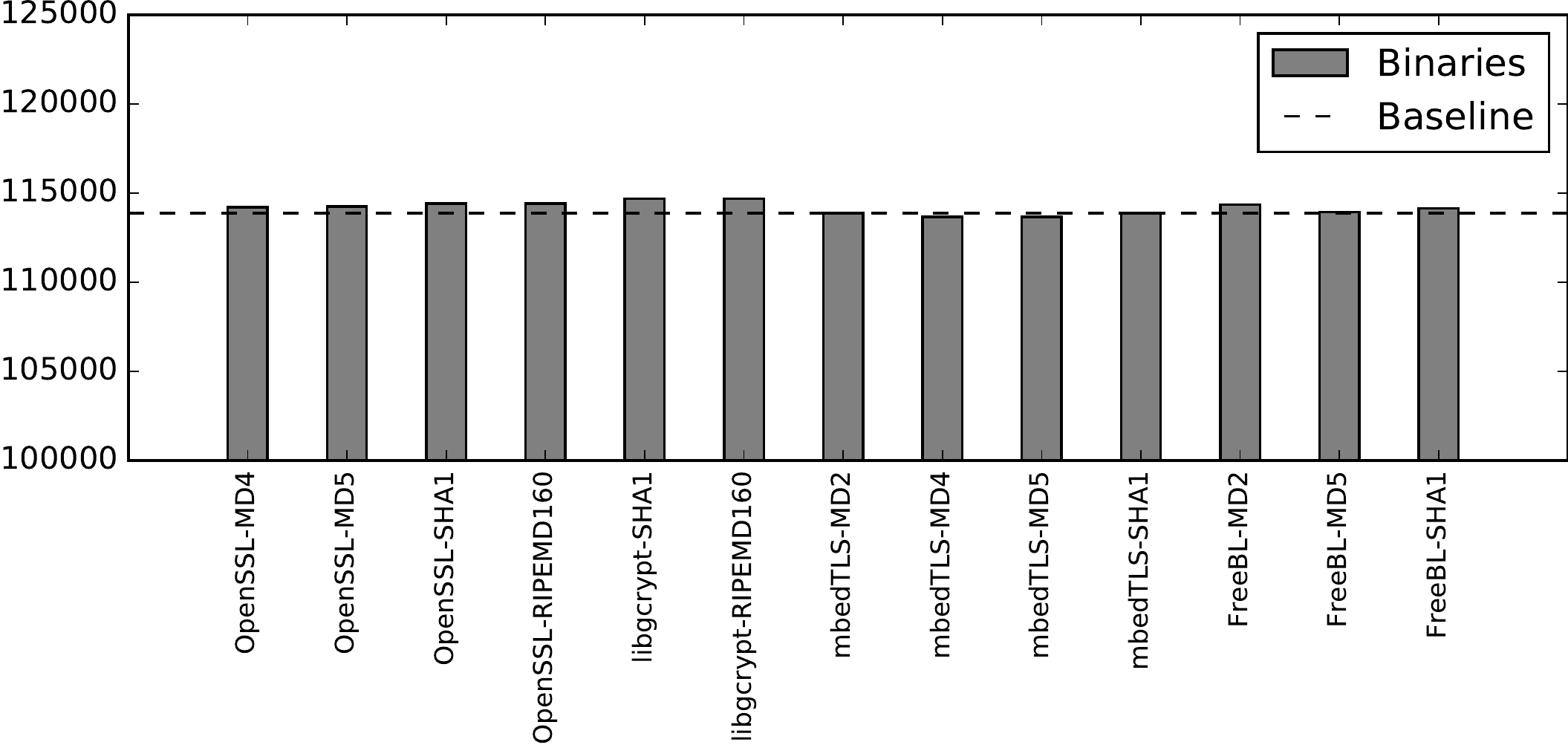}
  		\label{fig:ins-cryptolib}
  	}%
  	\subfloat[Runtime of \acron (in seconds).]{
  		\includegraphics[width=.5\linewidth]{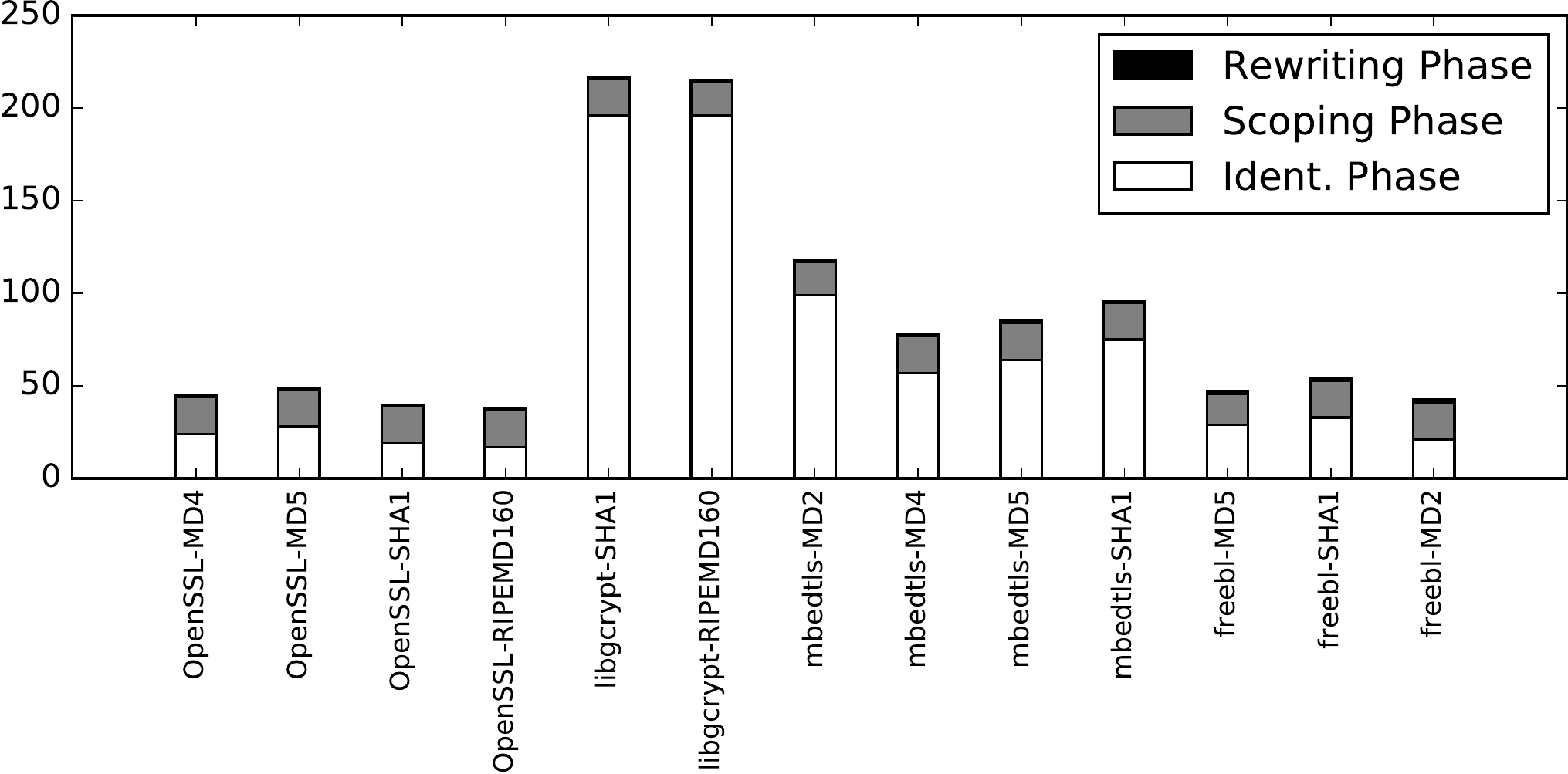}
  		\label{fig:runtime-cryptolib}
  	}
\caption{Evaluation results of \acron on cryptographic libraries. 
Original binaries are compiled with \texttt{O2}. Results for binaries with different optimization flags are similar and thus omitted.}
\label{fig:test}
\vspace{-.5cm}
\end{figure}

In terms of execution overhead, we implement a simple Pintool~\cite{pin} to count the number of instructions executed by the output binaries.
We then compare the result with the baseline, where manual editing is performed on the original binary's source code in order to replace the insecure hash function
and the modified source-code is properly optimized by the standard \texttt{gcc} compiler.
The results in Figure~\ref{fig:ins-cryptolib} show that the binaries produced by \acron have low execution overhead with an average of 300 added instructions, or only an increase of 0.3\%, compared to the baseline.
We also did not observe any noticeable increase in execution-time  for the output binaries.

\noindent\textbf{Toolchain Runtime.}
Figure~\ref{fig:runtime-cryptolib} shows runtime of \acron to produce the output binaries.
The total runtime heavily depends on the size of input binaries, and is dominated by the runtime of the identification phase.
This bottleneck happens because the identification phase involves heavy-weight analysis such as disassembling the entire binary and/or recovering the binary call graph.
It is also worth noting that such analysis is performed only once and \acron re-uses the analysis results in latter phases;
this leads to lower runtimes in the latter phases.

\subsection{Evaluation Results: Real-World Binaries}\label{sec:eval_rw}

Similar to the previous dataset, we manually inspect the binaries in this dataset to 
identify changes required in \fbox{\bf C3}, then supply them to \acron.
Such changes are mainly related to the digest size that is hard-coded in the source code.

\noindent\textbf{Correctness of Rewritten Binary.} \label{subsec:rwbin_correct}
\karim{We consider rewritten binaries to be correct if changes performed by \acron: (1) correctly implement new functionalities with respect to the target SHA-256 hash function and (2) do not interfere with the remaining functionalities. For instance, the former enforces the rewritten binary of \texttt{md5sum} to be able to perform \texttt{sha256sum} of a given input string. We realize the latter requirement by executing the binaries produced by \acron with all test cases (except the ones that use insecure hash functions) provided in their original respective project repository. 
We discuss expected functionalities of each test program in Appendix~\ref{apdx:exp_func}.}

\begin{table}[!t]
\footnotesize
\centering
\resizebox{.7\columnwidth}{!}{%
\begin{tabular}{|c|c|c|c|c||c|}
\hline
\multirow{2}{*}{\textbf{Program}} & \multirow{2}{*}{\textbf{OFLAG}} & \multicolumn{3}{c||}{\textbf{Binary Size}} & \textbf{Correctness of} \\ \cline{3-5} 
 &  & \textbf{O} & \textbf{R} & \textbf{$\Delta$} & \textbf{Output Binaries}\\ \hline
 \multirow{5}{*}{\texttt{md5sum}} & -O0 & 43.6KB & 51.9KB & 8.3KB & \cmark \\
 & -O1 & 35.4KB & 45.3KB & 9.9KB & \cmark \\
 & -O2 & 35.4KB & 45.0KB & 9.6KB & \cmark \\
 & -O3 & 39.5KB & 49.1KB & 9.6KB & \cmark \\
 & -Os & 31.3KB & 40.4KB & 9.1KB & \cmark \\ 
 \hline
 \multirow{5}{*}{\texttt{sha1sum}} & -O0 & 43.6KB & 51.9KB & 8.3KB & \cmark \\
 & -O1 & 35.4KB & 45.3KB & 9.9KB & \cmark \\
 & -O2 & 35.4KB & 45.0KB & 9.6KB & \cmark \\
 & -O3 & 39.5KB & 49.1KB & 9.6KB & \cmark \\
 & -Os & 31.3KB & 40.4KB & 9.1KB & \cmark \\ 
 \hline
 \multirow{5}{*}{\texttt{smd5\_mkpass}} & -O0 & 22.8KB & 29.9KB & 7.1KB & \cmark \\
 & -O1 & 18.7KB & 25.8KB & 7.1KB & \cmark \\
 & -O2 & 18.7KB & 25.8KB & 7.1KB & \cmark \\
 & -O3 & 22.8KB & 29.8KB & 7.0KB & \cmark \\
 & -Os & 18.7KB & 25.8KB & 7.1KB & \cmark \\ 
 \hline
 \multirow{5}{*}{\texttt{ssha\_mkpass}} & -O0 & 22.8KB & 29.9KB & 7.1KB & \cmark \\
 & -O1 & 18.7KB & 25.8KB & 7.1KB & \cmark \\
 & -O2 & 18.7KB & 25.8KB & 7.1KB & \cmark \\
 & -O3 & 22.8KB & 29.8KB & 7.0KB & \cmark \\
 & -Os & 18.7KB & 25.8KB & 7.1KB & \cmark \\ 
 \hline
 \multirow{5}{*}{\texttt{curl}} & -O0 & 929.6KB & 937.3KB & 7.7KB & \cmark \\
 & -O1 & 589.6KB & 596.1KB & 6.5KB & \cmark  \\
 & -O2 & 614.2KB & 620.7KB & 6.5KB & \cmark  \\
 & -O3 & 675.6KB & N/A & N/A & \xmark  \\
 & -Os & 528.1KB & 534.5KB & 6.4KB & \cmark \\ \hline
 \multirow{5}{*}{\texttt{lighttpd}} & -O0 & 720.2KB & 724.0KB & 3.8KB & \cmark \\
 & -O1 & 522.1KB & 529.1KB & 7.0KB & \cmark \\
 & -O2 & 534.7KB & 545.8KB & 11.2KB & \cmark \\
 & -O3 & 584.7KB & 592.0KB & 7.4KB & \cmark \\
 & -Os & 466.4KB & 473.1KB & 6.6KB & \cmark \\ 
 \hline
\end{tabular}
}
\caption{Size of original (O) and rewritten (R) binaries of real-world applications. $\Delta$ indicates the binary size overhead.}
\label{tab:alice-rwbin}
\vspace{-.7cm}
\end{table}

Table~\ref{tab:alice-rwbin} shows the correctness of output binaries produced by \acron in this dataset.
All output binaries pass all test cases in their original project repository
while only one output binary fails to pass the expected functionality.
We manually examined the failed binary and found that 
\acron misidentified an insecure hash routine.
Our further inspection reveals that the main culprit appears to be our underlying dynamic concrete executor, which
fails to output the expected MD5 digest even if \acron sets up a proper call stack.
As a result, \acron's identification phase did not detect this insecure hash function in the input binary,
and the output binary remained unchanged.

\begin{figure}
	\centering
  \includegraphics[width=.9\linewidth]{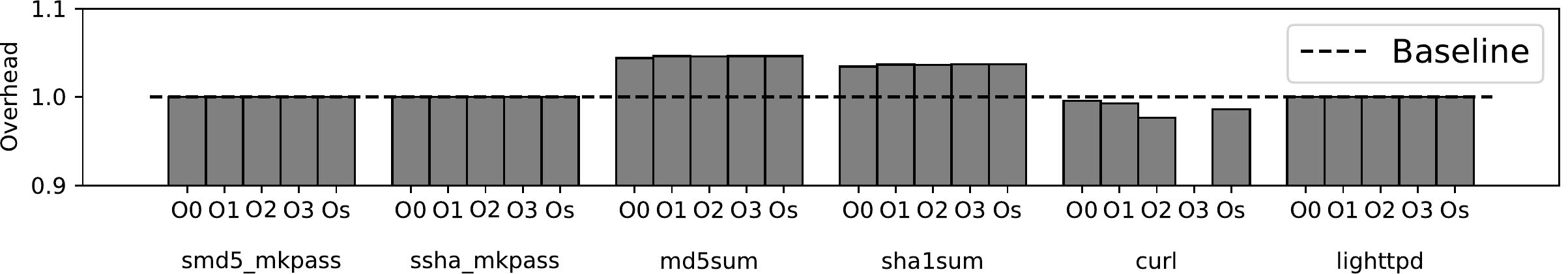}
  \caption{Overhead in terms of executed instructions of real-world binaries.}
  \label{fig:ins-rwbin}
\vspace{-.7cm}
\end{figure}

\noindent\textbf{Binary Size and Execution Overhead.}
Table~\ref{tab:alice-rwbin} shows the increase in binary size in this dataset. 
\acron adds around 4 to 11KB to the original binary. \karim{As mentioned in Section~\ref{sec:eval_lib}, up to 8KB of this overhead is caused by the underlying binary rewriter performing a patch alignment. The remaining overhead stems from rewritten functions that are appended at the end of the new binary.}
We note that we excluded the result for the \texttt{curl} binary compiled with \texttt{O3}
in Table~\ref{tab:alice-rwbin} (and subsequent figures) as \acron could not produce the correct output binary in that case.

We did not observe any noticeable execution overhead in terms of execution-time when running the output binary against the provided test cases from the project's repository.
In addition, we measure the number of executed instructions for the output binary to perform the expected functionality with respect to the SHA-256 function
and compare it to the baseline, where we manually edit the source code to replace the insecure hash function. 
The result, shown in Figure~\ref{fig:ins-rwbin}, also indicates negligible ($<5\%$) increase in execution-time in this case. \karim{Note that even though execution of the rewritten \texttt{curl} binaries becomes faster (requires 2\% fewer instructions), this improvement is still negligible. As such, we do not claim that \acron helps producing a more efficient output binary.}

\begin{figure}
	\centering
  \includegraphics[width=.9\linewidth]{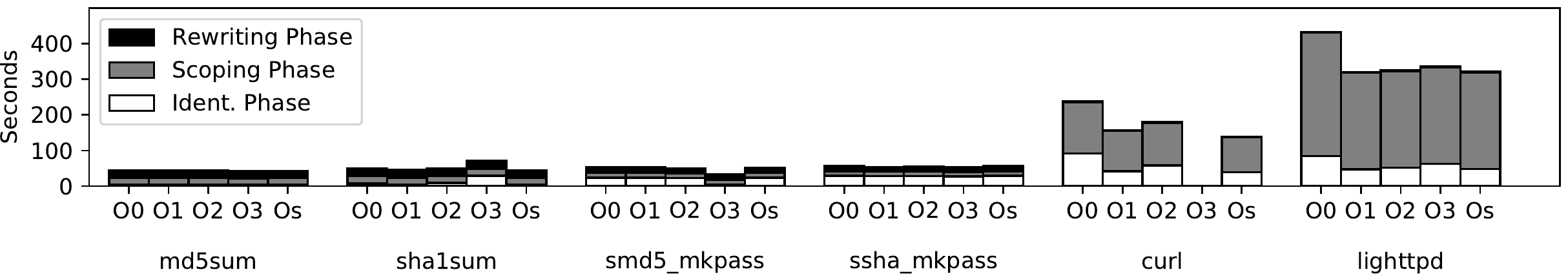}
  \caption{Runtime of the \acron toolchain on real-world binaries.}
  \label{fig:runtime-rwbin}
\end{figure}

\noindent\textbf{Toolchain Runtime.}
Figure~\ref{fig:runtime-rwbin} illustrates \acron's runtime to produce each output binary.
\karim{In simpler programs (e.g., \texttt{md5sum} or \texttt{smd5\_mkpass}), \acron identifies and replaces weak primitives in less than a minute.
For more complex programs (e.g., \texttt{lighttpd}), \acron's runtime can be a bit slower -- up to 5 minutes.
Most of the runtime overhead comes from the scoping phase
because it needs to instrument a large number of instructions (e.g, $\approx500$k instructions for \texttt{lighttpd}) while execution of simpler programs contains significantly fewer instructions.}
We consider \acron's runtime to be acceptable 
since the entire process only needs to be performed once, making 
the toolchain runtime not a primary concern.

\begin{table}[!h]
	\footnotesize
	\centering
	\resizebox{.8\textwidth}{!}{%
		\begin{tabular}{cc|cccccc||c} \hline
			\multicolumn{2}{c|}{\bf Program}            & \texttt{md5sum} & \texttt{sha1sum} & \texttt{smd5\_mkpass} & \texttt{ssha\_mkpass} & \texttt{curl}    & \texttt{lighttpd} & Avg. \\ \hline\hline
			\multirow{3}{*}{\bf Changes:}      & \fbox{\scriptsize\bf C1}      & 1024   & 1806    & 932          & 1676         & 424     & 1712     & 1262    \\
			& \fbox{\scriptsize\bf C2}      & 1      & 1       & 3            & 9            & 8       & 2        & 4       \\
			& \fbox{\scriptsize\bf C3}      & 1      & 1       & 2            & 3            & 1       & 2        & 1.7     \\ \hline
			\multicolumn{2}{c|}{\scriptsize(\fbox{\bf C1}+\fbox{\bf C2})/Total} & 99.9\% & 99.94\% & 99.79\%      & 99.85\%      & 99.78\% & 99.88\%  & 99.87\% \\ \hline
			\multicolumn{2}{c|}{\scriptsize\fbox{\bf C2}/(\fbox{\bf C2}+\fbox{\bf C3})}         & 50\%   & 50\%    & 60\%         & 75\%         & 87.5\%  & 50\%     & 70.17\% \\ \hline
		\end{tabular}%
	}
	\caption{Number of rewritten instructions required for each category of changes. All test programs are compiled with \texttt{O2} flag.}
	\label{tab:alice-quantify}
	\vspace{-.5cm}
\end{table}

\noindent\textbf{Reduction in Manual Efforts.} 
\karim{While \acron currently does not automatically identify changes from the \fbox{\bf C3} class, it still saves considerable manual effort.
We quantify such savings in Table~\ref{tab:alice-quantify} as the number of instructions required to be rewritten in order to implement changes for each category. On average, \acron automatically identifies and rewrites 1,266 instructions, which translates into 99.87\% reduction in manual efforts. However, we acknowledge that it may be possible to use existing cryptographic identification tools (with some modifications) to locate changes from \fbox{\bf C1}. Even when such tools exist, our toolchain still significantly reduces manual work by 70.17\%. It is worth emphasizing again that no existing tools are capable of identifying changes from \fbox{\bf C2} and \fbox{\bf C3}}.

\section{Limitations and Future Work}\label{sec:limitation}

The current version of \acron  has limitations. 
First, \acron relies on some underlying (open source) building-block tools and inherits their limitations.
For example, we encountered  instances where 
the underlying x86-64 assembler, \texttt{Keystone} \cite{keystone},
fails to translate uncommon instructions whose operands contain \texttt{fs} registers or a \texttt{rep} instruction. 
Whenever we encounter such an  issue, we manually fixed it by directly hard-coding the correct behavior into \acron.
Furthermore, the first phase of \acron relies on the \texttt{angr} framework~\cite{angr} for disassembly of stripped binaries;
\texttt{angr} does not perform static disassembly with correctness guarantees.
In fact, static disassembly of stripped binaries with correctness guarantees is still an open  problem~\cite{andriesse2016depth}.
Thus, \texttt{angr} may produce incorrect results in \acron's first phase, which affects outcomes of output binaries.

\acron also assumes that the routine implementing an insecure hash function and necessary changes
are statically included in the main application.  \acron does not currently support identifying and replacing insecure cryptographic primitives located in a dynamic library.
Expanding \acron's functionalities to dynamic libraries is possible since most of our underlying tools
are  capable of locating and analyzing  dynamic libraries used by the main application.

\acron does not automatically identify nor rewrite changes from \fbox{\bf C3} and relies on the user to supply them to the toolchain.
In practice, some manual effort is required to locate changes in binary logic for a large binary.
Automating this process is a challenging problem and is an interesting avenue for future work.
Due to space limitations, we discuss this issue in more detail in Appendix~\ref{apdx:limitation}.

Finally, we design \acron to target non-malicious legacy binaries and assume that such binaries are not obfuscated.
In practice, even legitimate software may make use of obfuscation techniques, e.g., to protect intellectual property.
Extending \acron to support obfuscated but non-malicious binaries (e.g., the software is not malware trying to evade analysis) 
 is also an interesting future direction.


\section*{Acknowledgments}
This work was sponsored by the U.S. Department of Homeland Security (DHS) Science and Technology (S\&T) Directorate under Contract No. HSHQDC-16-C-00034. Any opinions, findings, and conclusions or recommendations expressed in this material are those of the authors and do not necessarily reflect the views of DHS and should not be interpreted as necessarily representing the official policies or endorsements, either expressed or implied, of DHS or the U.S. government.
The authors thank the anonymous reviewers for their valuable comments.


\section{Conclusion  \label{conclusion}}

We have developed \acron, a toolchain for identifying weak or broken cryptographic primitives
and replacing them with secure ones without relying on source code or debugging symbols.
We have implemented a prototype of \acron that can detect several cryptographic primitives while only requiring access to the binaries containing them. 
Our implementation of \acron can also automatically replace weak and/or broken  implementations of cryptographic hash functions in ELF-based x86-64 binaries.
We have demonstrated \acron's effectiveness on various open-source cryptographic libraries and real-world applications utilizing cryptographic hash functions. 
Our experimental results show that \acron can successfully locate and replace insecure hash functions while preserving existing functionalities in the original binaries.

\bibliographystyle{IEEEtran}
\bibliography{biblio}

\newpage

\begin{subappendices}
\renewcommand{\thesection}{\Alph{section}}%

\section{Detectable Cryptographic Primitives}\label{apdx:detect_prim}

A list of cryptographic primitives that are detected by \acron is shown in Table~\ref{tab:detected-primitives}.

\begin{table}[h]
\resizebox{\linewidth}{!}{%
\begin{tabular}{|c|l|l|l|}
\hline
\textbf{Primitive} & \textbf{Name} & \textbf{Constant Type} & \textbf{Constant Values}  \\ \hline
 Cryptogrpahic & MD2 & S-Box & 292e43c9…  \\ \cline{2-4} 
  Hash& MD5 & IVs & {[}01234567, …, 76543210{]}  \\ \cline{2-4} 
 Functions& SHA1 & IVs & {[}01234567, …, f0e1d2c3{]}  \\ \cline{2-4} 
 & \begin{tabular}[c]{@{}l@{}}SHA-256\\ BLAKE2s\end{tabular} & IVs & {[}67e6096a, …, 19cde05b{]}  \\ \cline{2-4} 
 & \begin{tabular}[c]{@{}l@{}}SHA-512\\ BLAKE2b\end{tabular} & IVs & {[}08c9bcf367e6096a, …{]}  \\ \hline
 Block & RC2 & S-Box & d978f9c4…  \\ \cline{2-4} 
  Ciphers & RC5 & Magic Constant (P) & 6b2aed8a…  \\ \cline{2-4} 
   & Blowfish & S-Box & 886a3f24…  \\ \cline{2-4} 
  & DES & S-Box or SPtrans & 0e040d01… or 00080802…  \\ \cline{2-4} 
 & AES & S-Box & 637c777b…  \\ \hline
 Elliptic& NIST-P192 & Prime, Base Point x & {[}ffffffff…, 188da80e…{]} \\ \cline{2-4} 
 Curves   & NIST-P224 & Prime, Base Point x & {[}ffffffff…, b70e0cbd…{]}  \\ \cline{2-4} 
   (Signatures& NIST-P384 & Prime, Base Point x & {[}ffffffff…, aa87ca22…{]}  \\ \cline{2-4} 
  \& Key Exchange) & NIST-P256 & Prime, Base Point x & {[}ffffffff…, 6b17d1f2…{]}  \\ \cline{2-4} 
& NIST-P512 & Prime, Base Point x & {[}ffffffff…, b70e0cbd…{]}  \\ \hline
\end{tabular}%
}
\caption{A list of cryptographic primitives that can be detected by \acron. Note that only hash functions can also be automatically augmented in the current version of \acron.}
\label{tab:detected-primitives}
\end{table}

\section{Implementation Details of \acron \label{sec:impl}}

We describe below implementation details for each phase required in the operation of the \acron toolchain.\\

\noindent\textbf{Phase 1: Identifying and Locating Cryptographic Primitives.} We implement this phase 
on top of the binary analysis platform \texttt{angr}~\cite{angr}. 
Specifically, we use \texttt{angr}'s \texttt{CFGFast} API to perform the disassembly of instructions and generate the callgraph for a given binary.
\acron uses the resulting disassembly to determine parts of the binary containing specific constant vectors.
\acron analyzes the program callgraph to locate candidate routines.
Finally, to determine whether a candidate routine actually implements the target hash function,
we use \texttt{angr}'s built-in dynamic concrete executor to 
invoke each candidate routine on a given test string and compare the output with the expected output.
\\

\noindent\textbf{Phase 2: Scoping Changes.} \acron's second phase is built on top of a dynamic analysis framework, \texttt{Triton}~\cite{triton}.
In particular, we used \texttt{Triton}'s taint engine (which is implemented as a Pintool~\cite{pin}) for the implementation of our taint propagation rules.
\texttt{Triton} also provides an API for instrumenting a callback function to be executed at any point during the dynamic taint analysis.
We register two callback functions:

\begin{enumerate}
	\item The first callback is invoked immediately after the execution of each instruction. It implements our taint introduction rule
		by determining whether the current instruction: (1) is a call to, or a return from, the target hash routine;
		(2) results in any new tainted memory; (3) accesses static memory; and (4) changes a position of the stack frame.
	\item The second callback is executed at the end of dynamic taint analysis.
		It is responsible for aggregating all tainted memory cells into memory buffers and outputs the necessary results to the next phase.
\end{enumerate}

\begin{figure}

\begin{lstlisting}[style=CStyle]
void __attribute__ ((section(".ext_mem"))) size_t static_strlen(const char *str);

void __attribute__ ((section(".ext_mem"))) sha256(const BYTE data[], size_t len, BYTE hash[]);

void __attribute__ ((section(".ext_mem"))) sha256_init(SHA256_CTX *ctx);

void __attribute__ ((section(".ext_mem"))) sha256_update(SHA256_CTX *ctx, const BYTE data[], size_t len);

void __attribute__ ((section(".ext_mem"))) sha256_final(SHA256_CTX *ctx, BYTE hash[]);

void __attribute__ ((section(".in_len_out"))) in_len_out(const BYTE in[], size_t len, BYTE out[])
{
    sha256(in, len, out);
}

void __attribute__ ((section(".out_in"))) out_in(BYTE out[], const BYTE in[])
{
    size_t len = static_strlen(in);
    sha256(in, len, out);
}
\end{lstlisting}
\caption{Snippet of example source code for \texttt{SHA-256} patch.}
\label{fig:patch-c}
\end{figure}

\begin{figure}
\begin{lstlisting}[style=ShStyle]
#!/bin/sh
data_addr=$1
lib_addr=$2
entry_addr=$3
data_name=$4
lib_name=$5
entry_name=$6
gcc -g -O0 -fno-toplevel-reorder -fno-stack-protector -Wl,--section-start=.ext_mem=$lib_addr,--section-start=.ext_data=$data_addr,--section-start=.$entry_name=$entry_addr sha256.c -o sha256.o
objcopy --dump-section .ext_data=$data_name sha256.o
objcopy --dump-section .ext_mem=$lib_name sha256.o
objcopy --dump-section .$entry_name=$entry_name sha256.o
\end{lstlisting}
\caption{Example shell script used to generate a \texttt{SHA-256} patch.}
\label{fig:patch-sh}
\end{figure}

\noindent\textbf{Phase 3: Augmenting and Rewriting Changes.} 
There are two steps in this phase: (1) binary rewriting, and (2) patch generation.

First, we use a static binary rewriting tool, \texttt{patchkit}~\cite{patchkit}, to edit the input binary to support new changes.
After identifying and augmenting changes into the affected routines, 
we use \texttt{patchkit}'s \texttt{inject} API to add the modified routine into the new section of the original binary.
Then, we call the \texttt{hook} API to redirect all calls from the original routine to this modified routine.

Second, we generate a binary patch containing an implementation of the replacement hash routine.
Performing this is not trivial because we must ensure the following:
\begin{itemize}
	\item The binary patch must be self-contained, i.e., it must not rely on any external functions and/or libraries.
	To satisfy this, we statically included implementations of external functions used by the patch
	into its source code. Then, we create the patch based on such functions instead of the external functions/library.
	The example is shown in Line 1 of Figure~\ref{fig:patch-c} where the \texttt{glibc} \texttt{strlen()} is statically added into the patch's source code.
	
	\item Since the patch will be injected into a new section of a different binary, it must be compiled with respect to the address of this new section.
	The example shell script that we used to compile a \texttt{SHA-256} patch as well as the patch's source code are shown in Figure~\ref{fig:patch-sh} and~\ref{fig:patch-c}, respectively.
	At the source code level, we separate code and data of the patch by assigning them to different sections.
	Then, to ensure that they can be loaded from a specific address, we compile each with \texttt{--section-start} flag (Line 8 in Figure~\ref{fig:patch-sh}) and store the result into separate binaries (Line 9-10 in Figure~\ref{fig:patch-sh}).
	
	\item The hash function in the patch must have the same order of parameters as the target hash function.
	To achieve this, we include hash implementations of all possible parameter orders into the patch's source code
	and assigned each of the implementation to different binary sections (See Line 11-20 of Figure~\ref{fig:patch-c} for an example).
	Upon learning the parameters of the target hash function (from phase 2), we then select the section containing the corresponding implementation
	and integrate it into the patch at load-time (Line 11 in Figure~\ref{fig:patch-sh}).
\end{itemize}


\section{Recomputing Stack Offset}\label{apdx:alg}

In phase 3, once detecting a new stack frame size, \acron determines all instructions that access memory within the target stack frame.
If an access lies above any tainted buffer~\footnote{Recall that all tainted buffers are identified in \acron's phase 2.} (with respect to the \texttt{rsp} register),
\acron increments such access by $\delta$,
where $\delta$ is a sum of the size difference of all tainted buffers located below the given memory access.
Otherwise, the stack offset remains unchanged.
Details of this algorithm are shown in Algorithm~\ref{alg:stack-offset}.

\begin{figure}[!h]
  \centering
  \begin{minipage}{.65\linewidth}
\begin{algorithm}[H]\scriptsize
	\caption{Recompute stack offset}
	\label{alg:stack-offset}
    	\begin{algorithmic}[1]
		\Require $old\_offset$ - original stack offset w.r.t \texttt{rsp}
		\Require $buffers$ - a list of buffers that need to be expanded
        	\Ensure New stack offset
        	\Function {NewOffset}{$old\_offset, buffers$}
    			\State  $\delta = 0$
    			\ForAll {$b$ \textbf{in} $buffers$}
    				\If {$old\_offset < b.offset + b.old\_size$}
    					\State \textbf{break}
				\Else 
					\State $\delta = \delta + b.new\_size - b.old\_size$
				\EndIf
			\EndFor
		\State \textbf{return} $old\_offset + change$
        	\EndFunction
    	\end{algorithmic}
\end{algorithm}
  \end{minipage}
\end{figure}

\section{Locating Changes in Binary Logic}\label{apdx:limitation}

\begin{figure}[!hbt]
	\centering
	\begin{minipage}{.48\linewidth}
\lstset{language=[x64]Assembler,
	frame=single, 
        basicstyle=\tiny\ttfamily, 
        keywordstyle=\color{Blue}, 
	 morecomment=[l][\color{Orange}]{;},
        tabsize=5, 
        numbers=none, 
        firstnumber=1, 
        }
\begin{lstlisting}[linebackgroundcolor={\lstcolorlines[lgreen]{16}}]
<sprintf_loop_body>:
  4006ad:  movzx  eax,BYTE PTR [rbp+rax*1-0x50]	
  4006b2:  movzx  eax,al
  4006b5:  mov    edx,DWORD PTR [rbp-0x64]		
  4006b8:  add    edx,edx
  4006ba:  movsxd rdx,edx
  4006bd:  lea    rcx,[rbp-0x30]				
  4006c1:  add    rcx,rdx					
  4006c4:  mov    edx,eax
  4006c6:  mov    esi,0x4007d4
  4006cb:  mov    rdi,rcx
  4006ce:  mov    eax,0x0
  4006d3:  call   400500 <sprintf@plt>
<sprintf_loop_condition>:
  4006d8:  add    DWORD PTR [rbp-0x64],0x1		
  4006dc:  cmp    DWORD PTR [rbp-0x64],0xf
  4006e0:  jle    4006a8 <main+0x7f>
\end{lstlisting}
		\centering
		(a) Unoptimized Binary
	\end{minipage}
	\hfill
	\begin{minipage}{.48\linewidth}
\lstset{language=[x64]Assembler,
	frame=single, 
        basicstyle=\tiny\ttfamily, 
        keywordstyle=\color{Blue}, 
	 morecomment=[l][\color{Orange}]{;},
        tabsize=5, 
        numbers=none, 
        firstnumber=1, 
        }
\begin{lstlisting}[linebackgroundcolor={\lstcolorlines[lgreen]{2}}]
  400650:  lea    rbp,[rsp+0x10]
  400655:  lea    r14,[rsp+0x30]
  40065a:  mov    r12,rsp
  40065d:  mov    r13,0xffffffffffffffff
<sprintf_loop_body>:
  400664:  movzx  r8d,BYTE PTR [r12]
  400669:  mov    ecx,0x400784
  40066e:  mov    rdx,r13
  400671:  mov    esi,0x1
  400676:  mov    rdi,rbp
  400679:  mov    eax,0x0
  40067e:  call   4004f0 <__sprintf_chk@plt>
  400683:  add    r12,0x1
  400687:  add    rbp,0x2
<sprintf_loop_condition>:
  40068b:  cmp    rbp,r14
  40068e:  jne    400664 <main+0x5c>
\end{lstlisting}
		\centering
		(b) Optimized Binary
	\end{minipage}
	\caption{Partial disassembly of the \texttt{C} code from Figure~\ref{fig:crypto-example}a. Lines containing changes in binary logic are highlighted in green.}
	\label{fig:change_logic_exp}
\end{figure}

\acron currently does not automatically identify and rewrite changes in binary logic (\fbox{\bf C3}) but instead relies on the user to supply them to the toolchain.
In practice, some manual effort may be required to locate changes in binary logic for a large binary.
Automating this process without access to source code is currently a difficult open problem as we illustrate through a simple example below.

Figure~\ref{fig:change_logic_exp} shows the instructions corresponding to a loop (from Line 19 to 21) in the \texttt{C} code shown in Figure~\ref{fig:crypto-example}a compiled with and without optimizations.
Recall that operating changes in binary logic in this example corresponds to extending the loop termination conditions from the original digest size ($16$)
to the new digest size ($32$ in the case of SHA-256) at the source-code level.
Na\"ively, one could employ a heuristic based on the digest size to automatically identify such changes at the binary level:
by finding instructions containing the digest size value
and rewriting them with the new digest size value.
However, this heuristic would only be effective on the unoptimized binary
while it would fail on the optimized binary in Figure~\ref{fig:change_logic_exp}b.
This is because the loop in the optimized binary is transformed
in such a way that it terminates based on a different condition, i.e., the \texttt{hexdigest} buffer (at instruction \texttt{0x400655}) instead of the digest size value.
Therefore, solving this problem even for the simple program would require 
an approach more sophisticated than a simple heuristic. 
Solving this problem for general programs (e.g., the ones from Figure~\ref{fig:change-logic})
is more challenging.

\begin{figure}
\begin{minipage}[t]{0.48\textwidth}
\begin{lstlisting}[style=CStyle, escapechar=|, linebackgroundcolor={\lstcolorlines[lgreen]{8,10}},
        xleftmargin=0.1\linewidth,
        xrightmargin=0.1\linewidth
]
void f(char* input, ...)
{
	file_t *fp;
	size_t ds = 16;
	unsigned char d1[16], d2[16];
	...
	file_read(fp, d1, &ds);
	if(ds == 16 && error_check()) { |\label{line:16-first}|
		md5(d2, input, nbytes);
		if (memcmp(d1, d2, 16)) { |\label{line:16-second}|
			...
		}
	}
	...
}
\end{lstlisting}
{\raggedright\small
(a) Snippet of \texttt{mod\_slotmem\_shm.c} in \texttt{httpd-2.4.33}~\cite{httpd}: Change in binary logic includes editing 16 to 32.
This requires knowing that \texttt{16} in Line~\ref{line:16-first} and~\ref{line:16-second} refers to the size of the MD5 digest.
}
\end{minipage}\hfill
\begin{minipage}[t]{0.48\textwidth}
\begin{lstlisting}[style=CStyle,  linebackgroundcolor={\lstcolorlines[lgreen]{5}},
]
void f(...)
{
	...
	md5(folder, strlen (folder), &d);
	ret = snprintf(hcpath,_POSIX_PATH_MAX,"%s/%02x%02x%02x%02x%02x%02x%02x%02x%02x%02x%02x%02x%02x%02x%02x%02x-%s",path,d[0],d[1],d[2],d[3],d[4],d[5],d[6],d[7],d[8],d[9],d[10],d[11],d[12],d[13], d[14],d[15],chs);
	...
}
\end{lstlisting}
{\raggedright\small
(b) Snippet of \texttt{hcache.c} in \texttt{mutt-1.10}~\cite{mutt}: Change in binary logic includes (1): add \texttt{d[16...31]} as \texttt{snprintf()}'s parameters and
(2) modify the format string accordingly. This requires knowing that elements in \texttt{d} are passed into the function individually.}
\end{minipage}
\caption{Simplified snippets of two real-world programs that require operating changes in binary logic if \texttt{md5()} is replaced by \texttt{sha256()}.
Lines reflecting such changes at the source-code level are highlighted in green.}
\label{fig:change-logic}
\end{figure}

\karim{Given the difficulty of addressing \fbox{\bf C3} in its generality, it merits a separate treatment in another paper. 
Instead, we only outline here a semi-automated approach to address \fbox{\bf C3} that assumes access to source code and relies on minimal inputs from users. For example, assuming the user has access to the source code, they can identify obvious logic conditions that should be changed, e.g., loop terminating conditions that should be changed. For example, in Figure~\ref{fig:crypto-example}, replacing \texttt{MD5()} with \texttt{SHA-256()} by just enlarging \texttt{hexdigest} and \texttt{digest} variables does not suffice to produce the desired binary; one would have to also edit a loop terminating condition from $i < 16$ to $i < 32$ in line 19 and 24 to reflect the replacement SHA-256. We argue that it is reasonable to expect a user (which needs to only understand the interface of the SHA-256 hash function) to know that SHA-256 produces a digest with length of 32 bytes and that any code using such a digest should expect a buffer or variable of 32 bytes. The user can then instruct \acron that the terminating condition of the loop should be enlarged from 16 bytes to 32 bytes. We argue that a large portion of the \fbox{C3} class of changes can be addressed via such an approach, but leave it for future work to investigate this in more detail.}

\section{Expected Functionalities for Real-World Binaries}\label{apdx:exp_func}

This section discusses expected functionalities of both original and rewritten binaries for all real-world programs used in Section~\ref{sec:eval_rw}.
The expected functionalities of original programs are also used as test inputs in \acron's scoping phase in order to perform the dynamic taint analysis.
\ignore{
	To better illustrate correctness of the generated output binaries, 
	we first briefly describe our test applications and then the context in which insecure hash functions are used in them.}

\karim{
\texttt{md5sum} and \texttt{sha1sum} are popular checksum command line programs that can be used to compute the MD5 and SHA1 functions of a given file or input string. The expected functionality of the rewritten \texttt{md5sum} and \texttt{sha1sum} binaries is to be able to compute the SHA-256 function on a given input string and output it to the terminal.}

\karim{
The \texttt{smd5\_mkpass} program outputs a salted MD5 password from two inputs -- a user's secret key and a salt. This password then can be used to perform user authentication in the Lightweight Directory Access Protocol (LDAP)~\cite{ldap}. The \texttt{ssha\_mkpass} program works similarly but generates a salted SHA1 password instead of a salted MD5 password. Hence, we expect \acron to produce the rewritten \texttt{smd5\_mkpass} and \texttt{ssha\_mkpass} binaries capable of computing a salted SHA256 password from a secret key and a salt.}

\texttt{curl} is a client-side command-line utility tool for transferring data using URL syntax.
It supports various internet protocols such as HTTP(S), FTP(S), cookies, proxy tunneling or even
access authentication. 
In particular, we focus on evaluating \acron on \texttt{curl}'s implementation of digest access authentication (or \texttt{DigestAuth})
while removing other functionalities that use an insecure hash function in the \texttt{curl} binary.

\texttt{DigestAuth} provides an access control mechanism to a webserver. 
On the client side, \texttt{DigestAuth} constructs and sends the authorization token to the webserver by applying 
a hash function multiple times to the username, password and nonce generated by the webserver.
The webserver then makes a decision based on the received token whether to grant an access to requested resources or not.
Detailed description of \texttt{DigestAuth} is shown in Appendix~\ref{apdx:digest-auth}.

Particularly, version 7.56.0 of \texttt{curl} implements \texttt{DigestAuth}
based on the RFC 2617~\cite{rfc2617} that supports only MD5 as the underlying hash function.
Even though there is no known attack on such construction yet, it would still be a good idea to migrate
it to a more secure one.
Therefore, we expect \acron to produce the output \texttt{curl} binary capable of performing \texttt{DigestAuth} using SHA-256 instead of MD5.
Also, as part of this experiment, we implemented a test webserver in Python using Flask-HTTPAuth~\footnote{https://flask-httpauth.readthedocs.io}, which 
provides the \texttt{DigestAuth} functionality with MD5 or SHA-256 as the underlying hash function.

\texttt{lighttpd} provides a light-weight implementation of a webserver while remaining standards-compliant and flexible.
In particular, we are interested in evaluating \acron on \texttt{lighttpd}'s implementation of basic authentication (\texttt{BasicAuth}).
Similar to \texttt{DigestAuth}, \texttt{BasicAuth} is a mechanism to enforce access controls to web resources.
The main distinction is that a client in \texttt{BasicAuth} creates an authorization token by encoding 
the username and password with Base64. This is in contrast with a use of of a hash function in \texttt{DigestAuth}. 
The webserver then validates the authorization token by comparing it with the corresponding password stored in the password file.
To be secure, the password file must not store passwords in clear;
instead they should be stored in an encrypted or hashed format.
\texttt{lighttpd}'s \texttt{BasicAuth} supports various password encryption formats, including the insecure SHA1 hash function.
We provide further details of \texttt{BasicAuth} in Appendix~\ref{apdx:basic-auth}.
Our goal in this experiment is to replace SHA1 with SHA-256;
hence, after applying \acron on the \texttt{lighttpd} binary, we expect the output webserver binary to 
be able to perform \texttt{BasicAuth} using passwords stored as the SHA-256 format, instead of the SHA1 format.
For this experiment, we implemented a test webclient in Python to perform the client-side functionality of \texttt{BasicAuth}.

\ignore{
	For both programs, we must ensure that changes performed by \acron do not interfere with other existing functionalities
	besides the aforementioned functionalities.
	As such, we also ran the binaries produced by \acron against all test cases (except the ones that use insecure hash functions) provided in their original respective project repository.
	We then consider the output binary to be correct if it also passes all test cases.
	
	Lastly, recall that \acron's scoping phase requires executing the original binary on test inputs in order to perform the dynamic taint analysis.
	To this end, we created test inputs based on the original \texttt{DigestAuth} and \texttt{BasicAuth} schemes with insecure hash functions. 
}

\section{Details of Cryptographic Schemes}

\subsection{Digest Authentication Scheme}\label{apdx:digest-auth}

\begin{figure*}[!h]
	\centering
	\includegraphics[width=.7\linewidth]{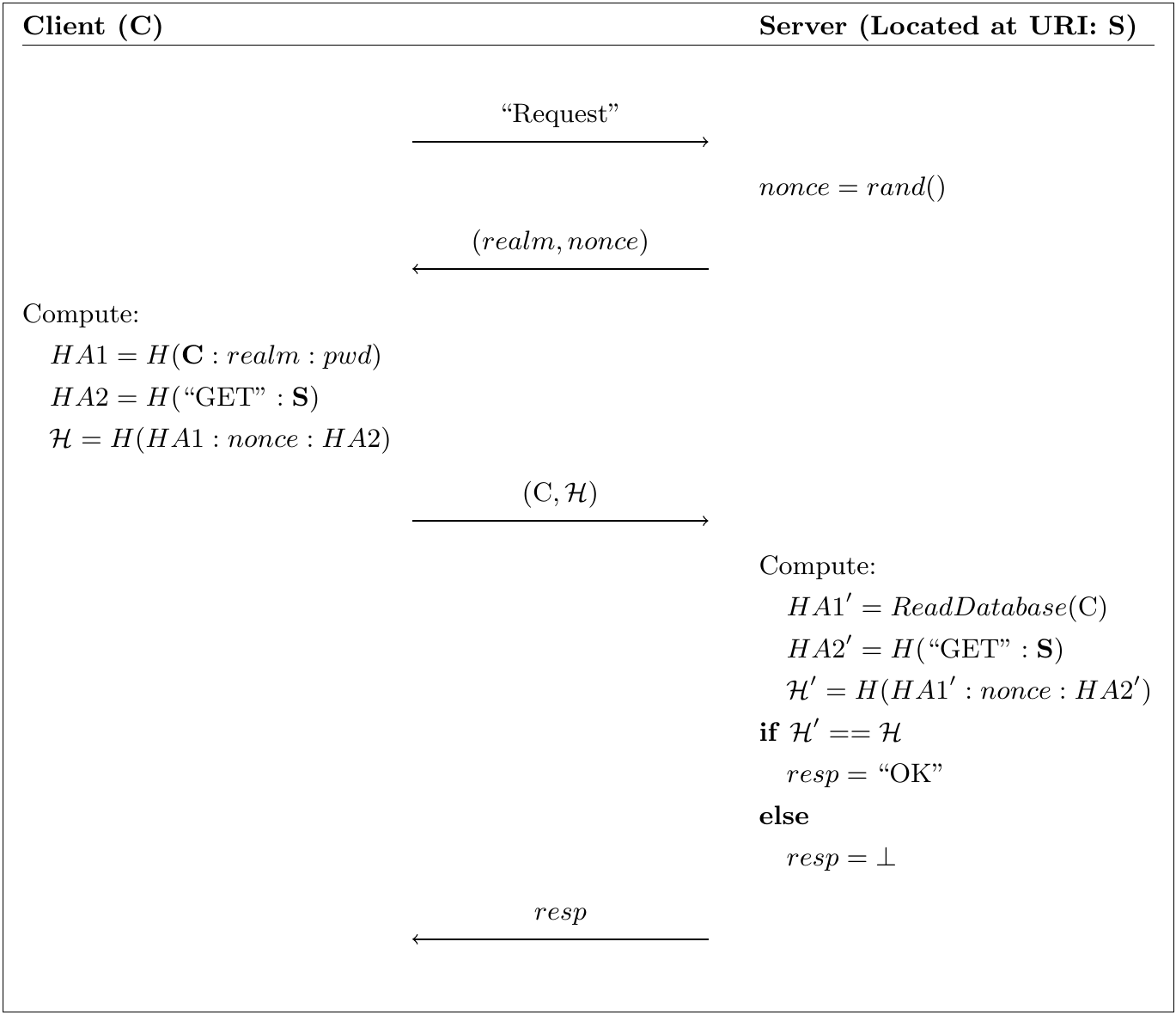}
	\caption{Digest Authentication Scheme}
	\label{fig:digest-auth}
\end{figure*}

Digest Authentication Scheme (\texttt{DigestAuth}) is depicted in details in Figure~\ref{fig:digest-auth}.
In our experiment from Section~\ref{sec:eval_rw}, the client communicates with the server through the \texttt{curl} executable
that implements \texttt{DigestAuth} using SHA1 as the underlying $H$ function.
The rewritten \texttt{curl} is considered correct if it can execute \texttt{DigestAuth} with $H(\cdot) \equiv SHA256(\cdot)$ and pass all of \texttt{curl}'s existing test cases.

\subsection{Basic Authentication Scheme}\label{apdx:basic-auth}

\begin{figure*}[!h]
	\centering
	\includegraphics[width=.7\linewidth]{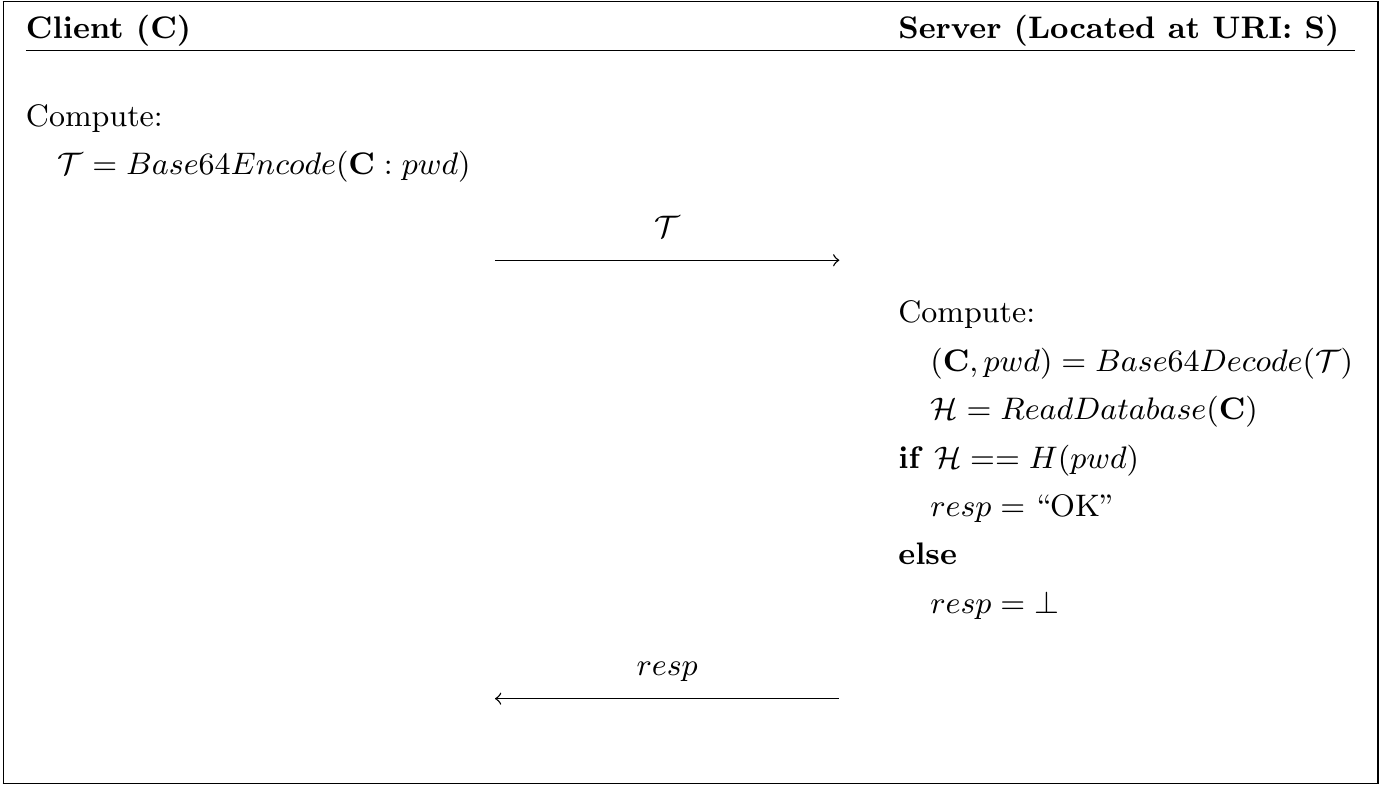}
	\caption{Basic Authentication Scheme}
	\label{fig:basic-auth}
\end{figure*}

Figure~\ref{fig:basic-auth} shows all steps required in Basic Authentication scheme (\texttt{BasicAuth}). 
In our experiment from Section~\ref{sec:eval_rw}, the server side is deployed using \texttt{lighttpd} and uses
MD5 as the underlying $H$ function.
Our goal is to locate the MD5 hash function in the \texttt{lighttpd} binary and replace it with SHA-256.
We consider the rewritten \texttt{lighttpd} produced by \acron to be correct if it can also follow the scheme in Figure~\ref{fig:basic-auth}
with $H(\cdot) \equiv SHA256(\cdot)$.

%


\section{Full Results for Size of Executables}\label{apdx:size_overhead}
We report increase in size of executable binaries produced by \acron on the cryptographic library dataset (see Section~\ref{sec:eval_lib}) in Table~\ref{tab:alice-cryptolib} below.
\begin{table}[!h]
\centering
\resizebox{\linewidth}{!}{%
\begin{tabular}{|c|c|c|c|c|c|}
\hline
\multirow{2}{*}{\textbf{Crypto Library}} & \multirow{2}{*}{\textbf{Algorithm}} & \multirow{2}{*}{\textbf{OFLAG}} & \multicolumn{3}{c|}{\textbf{Binary Size}} \\ \cline{4-6} 
 & \multicolumn{1}{l|}{} & \multicolumn{1}{l|}{} & O & R & $\Delta$ \\ \hline
\multirow{20}{*}{\shortstack{OpenSSL\\1.1.1}} & \multirow{5}{*}{MD4} & -O0 &  14.6KB  & 21.5KB  & 6.9KB  \\
 &  & -O1 & 14.6KB  & 21.5KB  & 6.9KB  \\
 &  & -O2 & 14.6KB & 21.4KB & 6.8KB  \\
 &  & -O3 & 14.6KB & 21.4KB & 6.8KB  \\
 &  & -Os & 14.6KB  & 21.3KB & 6.7KB \\ \cdashline{2-6} 
 & \multirow{5}{*}{MD5} & -O0 & 14.6KB  & 21.5KB &6.9KB  \\
 &  & -O1 & 14.6KB  & 21.5KB & 6.9KB  \\
 &  & -O2 &  14.6KB & 21.4KB  & 6.8KB  \\
 &  & -O3 & 14.6KB & 21.4KB  & 6.8KB  \\
 &  & -Os &  14.6KB &  21.3KB & 6.7KB  \\ \cdashline{2-6} 
 & \multirow{5}{*}{SHA1} & -O0 & 31.0KB & 37.8KB & 6.8KB   \\
 &  & -O1 & 31.0KB & 37.8KB & 6.8KB   \\
 &  & -O2 & 30.9KB & 37.8KB & 6.9KB  \\
 &  & -O3 & 30.9KB & 37.8KB & 6.9KB \\
 &  & -Os & 30.9KB & 37.7KB & 6.8KB  \\ \cdashline{2-6} 
 & \multirow{5}{*}{RIPEMD160} & -O0 & 22.8KB & 29.7KB & 6.9KB  \\
 &  & -O1 & 18.7KB & 25.5KB & 6.8KB \\
 &  & -O2 & 18.7KB & 25.5KB & 6.8KB  \\
 &  & -O3 & 18.7KB & 25.5KB & 6.8KB   \\
 &  & -Os & 18.6KB  & 25.4KB & 6.8KB  \\ \hline
 \multirow{15}{*}{\shortstack{FreeBL\\3.42}} & \multirow{5}{*}{MD2} & -O0 & 18.8KB & 25.6KB & 6.8KB  \\
 &  & -O1 & 18.8KB & 25.6KB & 6.8KB  \\
 &  & -O2 & 18.8KB & 25.6KB & 6.8KB   \\
 &  & -O3 & 22.9KB & 35.6KB & 12.7KB  \\
 &  & -Os & 18.8KB & 25.5KB& 6.7KB  \\ \cdashline{2-6} 
 & \multirow{5}{*}{MD5} & -O0 & 18.8KB & 28.1KB & 9.3KB \\
 &  & -O1 & 18.8KB & 28.1KKB & 9.3KB  \\
 &  & -O2 & 18.8KB & 25.5KB & 6.7KB \\
 &  & -O3 & 18.8KB & 25.5KB & 6.7KB  \\
 &  & -Os & 18.8KB & 25.4KB & 6.6KB  \\ \cdashline{2-6} 
 & \multirow{5}{*}{SHA1} & -O0 & 27.0KB & 40.4KB & 13.4KB \\
 &  & -O1 & 22.9KB & 32.2KB & 9.3KB  \\
 &  & -O2 & 22.9KB & 29.6KB & 6.7KB  \\
 &  & -O3 & 22.9KB & 29.6KB & 6.7KB  \\
 &  & -Os & 22.9KB & 29.5KB & 6.6KB  \\ \hline
\end{tabular}%
\quad
\begin{tabular}{|c|c|c|c|c|c|}
	\hline
	\multirow{2}{*}{\textbf{Crypto Library}} & \multirow{2}{*}{\textbf{Algorithm}} & \multirow{2}{*}{\textbf{OFLAG}} & \multicolumn{3}{c|}{\textbf{Binary Size}} \\ \cline{4-6} 
	& \multicolumn{1}{l|}{} & \multicolumn{1}{l|}{} & O & R & $\Delta$ \\ \hline
\multirow{20}{*}{\shortstack{mbedTLS\\2.16.0}} & \multirow{5}{*}{MD2} & -O0 & 10.5KB & 17.4KB & 6.9KB  \\
 &  & -O1 & 10.5KB & 17.4KB & 6.9KB \\
 &  & -O2 & 10.5KB & 17.3KB & 6.8KB \\
 &  & -O3 & 14.6KB & 28.0KB & 13.4KB \\
 &  & -Os & 10.5KB & 17.3KB & 6.8KB  \\ \cdashline{2-6} 
 & \multirow{5}{*}{MD4} & -O0 & 14.6KB & 24.0KB & 9.4KB  \\
 &  & -O1 & 14.6KB & 24.0KB & 9.4KB \\
 &  & -O2 & 10.5KB & 21.4KB & 10.9KB \\
 &  & -O3 & 14.6KB & 23.9KB & 9.3KB  \\
 &  & -Os & 10.5KB & 17.3KB & 6.8KB \\ \cdashline{2-6} 
 & \multirow{5}{*}{MD5} & -O0 & 14.6KB & 24.0KB & 9.4KB \\
 &  & -O1 & 14.6KB & 24.0KB & 9.4KB  \\
 &  & -O2 & 14.6KB & 21.4KB & 6.8KB  \\
 &  & -O3 & 14.6KB & 23.9KB & 9.3KB \\
 &  & -Os & 10.5KB & 17.3KB & 6.8KB \\ \cdashline{2-6} 
 & \multirow{5}{*}{SHA1} & -O0 & 18.7KB & 28.1KB & 9.4KB \\
 &  & -O1 & 14.6KB & 24.0KB & 9.4KB \\
 &  & -O2 & 14.6KB & 21.4KB & 6.8KB \\
 &  & -O3 & 18.7KB & 28.0KB & 9.3KB  \\
 &  & -Os & 14.6KB & 21.4KB & 6.8KB \\ \hline
 \multirow{10}{*}{\shortstack{libgcrypt\\1.8.4}} & \multirow{5}{*}{SHA1} & -O0 & 1487KB & 1492KB & 5KB  \\
 &  & -O1 & 1186KB & 1189KB & 3KB \\
 &  & -O2 & 1206KB & 1209KB & 3KB \\
 &  & -O3 & 1354KB & 1365KB & 11KB \\
 &  & -Os & 1120KB & 1123KB & 3KB  \\ \cdashline{2-6} 
 & \multirow{5}{*}{RIPEMD160} & -O0 & 1487kB & 1492KB & 5KB \\
 &  & -O1 & 1186KB & 1189KB & 3KB \\
 &  & -O2 & 1206KB & 1209KB & 3KB \\
 &  & -O3 & 1354KB & 1365KB & 11KB \\
 &  & -Os & 1120KB & 1123KB & 3KB \\ \hline
\end{tabular}%
}
\caption{Size of original (O) and rewritten (R) binaries created by compiling the application in Figure~\ref{fig:crypto-example}a that statically links to various crypto libraries with different optimization flags. $\Delta$ indicates the binary size overhead.}
\label{tab:alice-cryptolib}
\end{table}

\end{subappendices}

\end{document}